\documentclass[longauth]{aa}

\usepackage{txfonts}
\usepackage{graphicx}
\usepackage{natbib}
\usepackage{xspace}
\bibpunct{(}{)}{;}{a}{}{,}

\def\mhmpc{\,h^{-1}~\rm{Mpc}}

\def\mhmpcc{\,h^{-3}~\rm{Mpc^3}}
\def\mhgpcc{\,h^{-3}~\rm{Gpc^3}}

\def\mhmsun{\,h^{-1}~\rm{M_{\odot}}}

\def\Pdd{P_{\rm{\delta\delta}}}
\def\Pdt{P_{\rm{\delta\theta}}}
\def\Ptt{P_{\rm{\theta\theta}}}

\newcommand{\wprp}{$\,w_p(r_p)$\xspace}
\newcommand{\xisp}{$\,\xi(r_p,\pi)$\xspace}
\newcommand{\hmpc}{$\,h^{-1}$ Mpc\xspace}
\newcommand{\kms}{$\,{\rm km\, s^{-1}}$\xspace}
\newcommand{\SSR}{${SSR}$\xspace}
\newcommand{\TSR}{${TSR}$\xspace}
\newcommand{\CSR}{${CSR}$\xspace}
\newcommand{\ESR}{${ESR}$\xspace}

\begin{document}

\title{The VIMOS Public Extragalactic Redshift Survey (VIPERS)\thanks{
    based on observations collected at the European Southern
    Observatory, Cerro Paranal, Chile, using the Very Large Telescope
    under programs 182.A-0886 and partly 070.A-9007.  Also based on
    observations obtained with MegaPrime/MegaCam, a joint project of
    CFHT and CEA/DAPNIA, at the Canada-France-Hawaii Telescope (CFHT),
    which is operated by the National Research Council (NRC) of
    Canada, the Institut National des Sciences de l’Univers of the
    Centre National de la Recherche Scientifique (CNRS) of France, and
    the University of Hawaii. This work is based in part on data
    products produced at TERAPIX and the Canadian Astronomy Data
    Centre as part of the Canada-France-Hawaii Telescope Legacy
    Survey, a collaborative project of NRC and CNRS. The VIPERS web
    site is http://www.vipers.inaf.it/.}  }

\subtitle{Galaxy clustering and redshift-space distortions at
  $\mathbf{z\simeq0.8}$ in the first data release}

\titlerunning{Galaxy clustering and redshift-space distortions in VIPERS}

\author{
S.~de~la~Torre\inst{1}
\and L.~Guzzo\inst{2,3}
\and J.~A.~Peacock\inst{1}
\and E.~Branchini\inst{4,5,6}
\and A.~Iovino\inst{2}
\and B.~R.~Granett\inst{2}
\and U.~Abbas\inst{7}
\and C.~Adami\inst{8}
\and S.~Arnouts\inst{9,8}
\and J.~Bel\inst{10}
\and M.~Bolzonella\inst{11}
\and D.~Bottini\inst{12}
\and A.~Cappi\inst{11,13}
\and J.~Coupon\inst{14}
\and O.~Cucciati\inst{11}
\and I.~Davidzon\inst{11,15}
\and G.~De Lucia\inst{16}
\and A.~Fritz\inst{12}
\and P.~Franzetti\inst{12}
\and M.~Fumana\inst{12}
\and B.~Garilli\inst{12,8}
\and O.~Ilbert\inst{8}
\and J.~Krywult\inst{17}
\and V.~Le Brun\inst{8}
\and O.~Le F\`evre\inst{8}
\and D.~Maccagni\inst{12}
\and K.~Ma{\l}ek\inst{18}
\and F.~Marulli\inst{15,11,19}
\and H.~J.~McCracken\inst{20}
\and L.~Moscardini\inst{15,11,19}
\and L.~Paioro\inst{12}
\and W.~J.~Percival\inst{21}
\and M.~Polletta\inst{12}
\and A.~Pollo\inst{22,23}
\and H.~Schlagenhaufer\inst{24,25}
\and M.~Scodeggio\inst{12}
\and L.~A.~.M.~Tasca\inst{8}
\and R.~Tojeiro\inst{21}
\and D.~Vergani\inst{26}
\and A.~Zanichelli\inst{27}
\and A.~Burden\inst{21}
\and C.~Di Porto\inst{11}
\and A.~Marchetti\inst{28,2}
\and C.~Marinoni\inst{10}
\and Y.~Mellier\inst{20}
\and P.~Monaco\inst{29,16}
\and R.~C.~Nichol\inst{21}
\and S.~Phleps\inst{25}
\and M.~Wolk\inst{20}
\and G.~Zamorani\inst{11}
}

\institute{
SUPA, Institute for Astronomy, University of Edinburgh, Royal Observatory, Blackford Hill, Edinburgh EH9 3HJ, UK %14
\and INAF - Osservatorio Astronomico di Brera, Via Brera 28, 20122 Milano, via E. Bianchi 46, 23807 Merate, Italy %2
\and Dipartimento di Fisica, Universit\`a di Milano-Bicocca, P.zza della Scienza 3, I-20126 Milano, Italy %27
\and Dipartimento di Matematica e Fisica, Universit\`{a} degli Studi Roma Tre, via della Vasca Navale 84, 00146 Roma, Italy %10
\and INFN, Sezione di Roma Tre, via della Vasca Navale 84, I-00146 Roma, Italy %28
\and INAF - Osservatorio Astronomico di Roma, via Frascati 33, I-00040 Monte Porzio Catone (RM), Italy %29
\and INAF - Osservatorio Astrofisico di Torino, 10025 Pino Torinese, Italy %5
\and Aix Marseille Universit\'e, CNRS, LAM (Laboratoire d'Astrophysique de Marseille) UMR 7326, 13388, Marseille, France  %4
\and Canada-France-Hawaii Telescope, 65--1238 Mamalahoa Highway, Kamuela, HI 96743, USA %6
\and Centre de Physique Th\'eorique, UMR 6207 CNRS-Universit\'e de Provence, Case 907, F-13288 Marseille, France %7
\and INAF - Osservatorio Astronomico di Bologna, via Ranzani 1, I-40127, Bologna, Italy %9
\and INAF - Istituto di Astrofisica Spaziale e Fisica Cosmica Milano, via Bassini 15, 20133 Milano, Italy%3
\and Laboratoire Lagrange, UMR7293, Universit\'e de Nice Sophia-Antipolis, CNRS, Observatoire de la C\^ote d'Azur, 06300 Nice, France %31
\and Institute of Astronomy and Astrophysics, Academia Sinica, P.O. Box 23-141, Taipei 10617, Taiwan%12
\and Dipartimento di Fisica e Astronomia - Universit\`{a} di Bologna, viale Berti Pichat 6/2, I-40127 Bologna, Italy %17
\and INAF - Osservatorio Astronomico di Trieste, via G. B. Tiepolo 11, 34143 Trieste, Italy %13
\and Institute of Physics, Jan Kochanowski University, ul. Swietokrzyska 15, 25-406 Kielce, Poland %15
\and Department of Particle and Astrophysical Science, Nagoya University, Furo-cho, Chikusa-ku, 464-8602 Nagoya, Japan %16
\and INFN, Sezione di Bologna, viale Berti Pichat 6/2, I-40127 Bologna, Italy %18
\and Institute d'Astrophysique de Paris, UMR7095 CNRS, Universit\'{e} Pierre et Marie Curie, 98 bis Boulevard Arago, 75014 Paris, France %19
\and Institute of Cosmology and Gravitation, Dennis Sciama Building, University of Portsmouth, Burnaby Road, Portsmouth, PO1 3FX %11
\and Astronomical Observatory of the Jagiellonian University, Orla 171, 30-001 Cracow, Poland %22
\and National Centre for Nuclear Research, ul. Hoza 69, 00-681 Warszawa, Poland %23
\and Universit\"{a}tssternwarte M\"{u}nchen, Ludwig-Maximillians Universit\"{a}t, Scheinerstr. 1, D-81679 M\"{u}nchen, Germany %24
\and Max-Planck-Institut f\"{u}r Extraterrestrische Physik, D-84571 Garching b. M\"{u}nchen, Germany %20
\and INAF - Istituto di Astrofisica Spaziale e Fisica Cosmica Bologna, via Gobetti 101, I-40129 Bologna, Italy %25
\and INAF - Istituto di Radioastronomia, via Gobetti 101, I-40129, Bologna, Italy %26
\and Universit\`{a} degli Studi di Milano, via G. Celoria 16, 20130 Milano, Italy %1
\and Dipartimento di Fisica dell'Universit\`a di Trieste, Sezione di Astronomia, Via Tiepolo 11, I-34131 Trieste, Italy %30
}

\authorrunning{S. de la Torre et al.}

\offprints{\mbox{S.~de~la~Torre}, \email{sdlt@roe.ac.uk}}

\abstract{We present in this paper the general real- and
  redshift-space clustering properties of galaxies as measured in the
  first data release of the VIPERS survey. VIPERS is a large redshift
  survey designed to probe in detail the distant Universe and its
  large-scale structure at $0.5<z<1.2$. We describe in this analysis
  the global properties of the sample and discuss the survey
  completeness and associated corrections. This sample allows us to
  measure the galaxy clustering with an unprecedented accuracy at
  these redshifts. From the redshift-space distortions observed in the
  galaxy clustering pattern we provide a first measurement of the
  growth rate of structure at $z=0.8$: $f\sigma_8=0.47\pm 0.08$. This
  is completely consistent with the predictions of standard
  cosmological models based on Einstein gravity, although this
  measurement alone does not discriminate between different gravity
  models.}


\maketitle

\section{Introduction}

Over the past decades galaxy redshift surveys have provided a wealth
of information on the inhomogeneous universe, mapping the late-time
development of the small metric fluctuations that existed at early
times, and whose early properties can be viewed in the cosmic
microwave background (CMB).  The growth of structure during this
intervening period is sensitive both to the type and amount of dark
matter, and also to the theory of gravity, so there is a strong
motivation to make precise measurements of the rate of growth of
cosmological structure \citep[e.g.][]{jain2010}.

Of course, galaxy surveys do not image the mass fluctuations directly,
unlike gravitational lensing. But the visible light distribution does
have some advantages as a cosmological tool in comparison with
lensing. The number density of galaxies is sufficiently high that the
density field of luminous matter can be measured with a finer spatial
resolution, probing interesting nonlinear features of the clustering
pattern with good signal-to-noise. The price to be paid for this is
that the complicated biasing relation between visible and dark matter
has to be confronted; but this is a positive factor in some ways,
since understanding galaxy formation is one of the main questions in
cosmology.  Redshift surveys provide the key information needed to
meet this challenge: global properties of the galaxy population and
their variation with environment and with epoch.

The final advantage of redshift surveys is that the radial information
depends on cosmological expansion and is corrupted by peculiar
velocities. Although the lack of a simple method to recover true
distances can be frustrating at times, it has come to be appreciated
that this complication is in fact a good thing. The peculiar
velocities induce an anisotropy in the apparent clustering, from which
the properties of the peculiar velocities can be inferred much more
precisely than in any attempt to measure them directly using distance
estimators.  The reason peculiar velocities are important is that they
are related to the underlying linear fractional density perturbation
$\delta$ via the continuity equation: $\dot\delta =
-{\mathbf{\nabla\cdot u}}$, where $\vec{u}$ is the peculiar velocity
field. This can be expressed more conveniently in terms of the
dimensionless scale factor, $a(t)$, and the Hubble parameter, $H(t)$,
as
\begin{equation}
{\vec{\nabla\cdot u}}= -H f \delta; \quad f\equiv {d\ln \delta\over d\ln a}.
\end{equation}
The growth rate can be approximated in most models by $f(a)\simeq
\Omega_m(a)^\gamma$, where $\gamma\simeq 0.545$ in standard
$\Lambda$-dominated models, but where models of non-standard gravity
display a growth rate in which the effective value of $\gamma$ can
differ by $30\%$ \citep{linder2007}.

The possibility of using the redshift-space distortion signature as a
probe of the growth rate of density fluctuations, together with
  that of using the Baryonic Acoustic Oscillations (BAO) as a standard
  ruler to measure the expansion history, is one of the main reasons
behind the recent burst of activity in galaxy redshift surveys. The
first paper to emphasise this application as a test of gravity
theories was the analysis of the VVDS survey by \citet{guzzo08}, and
subsequent work especially by the SDSS LRG \citep{samushia12}, WiggleZ
\citep{blake12,contreras13}, 6dFGS \citep{beutler12} and BOSS
\citep{reid12} surveys has exploited this method to make measurements
of the growth rate at $z<1$. 

Surveys such as SDSS LRG, WiggleZ, or BOSS are characterized by a
large volume ($0.5-2\, h^{-3}{\rm Gpc}^3$), and a relatively sparse
galaxy population with number density of about $10^{-4}\,h^{3}{\rm
  Mpc}^{-3}$.  Statistical errors are in this case minimized thanks to
the large volume probed, at the expenses of selecting a very specific
galaxy population (e.g. blue star forming or very massive galaxies),
often with a complex selection function.  The goal of the VIMOS Public
Extragalactic Redshift Survey (VIPERS, {\tt http://vipers.inaf.it}),
has been that of constructing a survey with broader science goals and
properties comparable to local general-purpose surveys such as the
2dFGRS.  The adopted strategy has been to optimise the features of the
ESO VLT multi-object spectrograph VIMOS in order to measure about
$\sim 400$ spectra at $I_{AB}<22.5$ over an area of $\sim 200$ square
arcmin, in a single exposure of less than $1$ hour. The survey is
being performed as a ``Large Programme'' within the ESO general user
framework and aims at measuring redshifts for about $10^5$ galaxies at
$0.5<z<1.2$.

The prime goal of VIPERS is an accurate measurement of the growth rate
of large-scale structure at redshift around unity. The survey should
enable us in particular to use of techniques aimed at improving the
precision on the growth rate \citep{mcdonald09} thanks to its high
galaxies sampling of about $10^{-2}\,h^{3}{\rm Mpc}^{-3}$.  In
general, VIPERS is intended to provide robust and precise measurements
of the properties of the galaxy population at an epoch when the
Universe was about half its current age, representing one of the
largest spectroscopic surveys of galaxies ever conducted at these
redshifts. Examples can be found in the parallel papers that are part
of the first science release \citep{marulli13,malek13,davidzon13}.

This paper presents the initial analysis of the real-space galaxy
clustering and redshift-space distortions in VIPERS, together with the
resulting implications for the growth rate. The data are described in
Section 2; Section 3 describes the survey selection effects; Section 4
describes our methods for estimating clustering, which are tested on
simulations in Section 5; Section 6 presents the real-space clustering
results; Section 7 gives the redshift-space distortions results, and
Section 8 summarises our results and concludes.

Throughout this analysis, if not specified otherwise, we assume a
fiducial $\Lambda {\rm CDM}$ cosmological model with
$(\Omega_m,\Omega_k,w,\sigma_8,n_s)=(0.25,0,-1,0.8,0.95)$ and a Hubble
constant of $H_0=100~h~\rm{km~s^{-1}~Mpc^{-1}}$.

\section{Data}

The VIPERS galaxy target sample is selected from the optical
photometric catalogues of the Canada-France-Hawaii Telescope Legacy
Survey Wide \citep[CFHTLS-Wide,][]{goranova09}.  VIPERS covers $24$
deg$^2$ on the sky, divided over two areas within the W1 and W4 CFHTLS
fields. Galaxies are selected to a limit of $i'_{AB}<22.5$, applying a
simple and robust $gri$ colour pre-selection to efficiently remove
galaxies at $z<0.5$. Coupled with a highly optimized observing
strategy \citep{scodeggio09}, this allows us to double the galaxy
sampling rate in the redshift range of interest, with respect to a
pure magnitude-limited sample.  At the same time, the area and depth
of the survey result in a relatively large volume, $5 \times
10^{7}\mhmpcc$, analogous to that of the Two Degree Field Galaxy
Redshift Survey (2dFGRS) at $z\simeq0.1$ \citep{colless01,colless03}.
Such a combination of sampling rate and depth is unique amongst
current redshift surveys at $z>0.5$. VIPERS spectra are collected with
the VIMOS multi-object spectrograph \citep{lefevre03} at moderate
resolution ($R=210$) using the LR Red grism, providing a wavelength
coverage of 5500-9500$\rm{\AA}$ and a typical radial velocity error of
$\sigma_v=175(1+z)$\kms. The full VIPERS area of $24$ deg$^2$ will be
covered through a mosaic of 288 VIMOS pointings (192 in the W1 area,
and 96 in the W4 area).  A discussion of the survey data reduction and
management infrastructure is presented in \citet{garilli12}.  An early
subset of the spectra used here is analysed and classified through a
Principal Component Analysis (PCA) in \citet{marchetti13}.  A complete
description of the survey construction, from the definition of the
target sample to the actual spectra and redshift measurements, is
given in the parallel survey description paper \citep{guzzo13}.

The data set used in this and the other papers of the early science
release, will represent the VIPERS Public Data Release 1 (PDR-1)
catalogue. It will be publicly available in the fall of 2013.  This
catalogue includes $55358$ redshifts ($27935$ in W1 and $27423$ in W4)
and corresponds to the reduced data frozen in the VIPERS database at
the end of the 2011/2012 observing campaign; this represents $64\%$ of
the final survey in terms of covered area. A quality flag has been
assigned to each object in the process of determining their redshift
from the spectrum, which quantifies the reliability of the measured
redshifts. In this analysis, we use only galaxies with flags 2 to 9
inclusive, corresponding to a sample with a redshift confirmation rate
of $98\%$. The redshift confirmation rate and redshift accuracy have
been estimated using repeated spectroscopic observations in the VIPERS
fields \citep[see][for details]{guzzo13}. The catalogue, which we will
refer to just as the VIPERS sample in the following, corresponds to a
sub-sample of $45871$ galaxies with reliable redshift measurements.

\begin{figure}
\resizebox{\hsize}{!}{\includegraphics{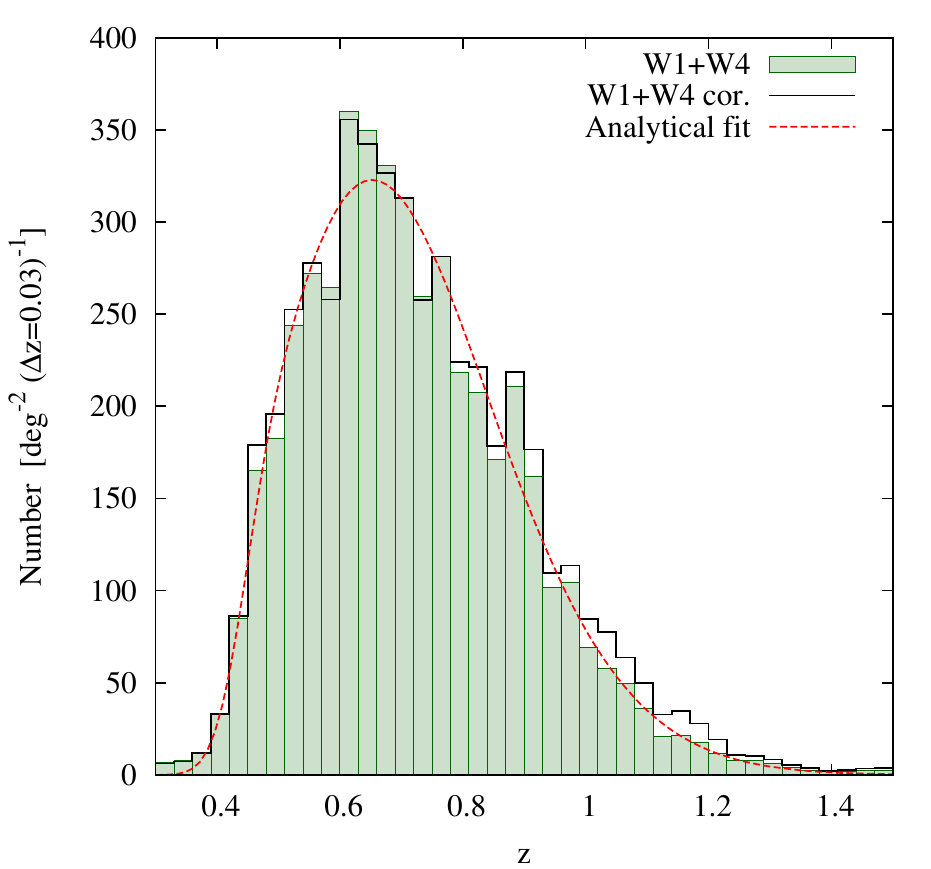}}
\caption{Redshift distribution of the combined W1+W4 galaxy sample
  when including only reliable redshifts (filled histogram) and that
  corrected for the full survey completeness (empty histogram) scaled
  down by 40\% (see text). The curve shows the best-fitting template
  redshift distribution given by Eq. \ref{eq:nz} applied to the
  uncorrected observed distribution.}
\label{fig1}
\end{figure}

The redshift distribution of the sample is presented in
Fig. \ref{fig1}. We can see in this figure that the survey colour
selection allows an efficient removal of galaxies below $z=0.5$. It is
important to notice that the colour selection does not introduce a
sharp cut in redshift but a redshift window function which has a
smooth transition from zero to one in the redshift range $0.4<z<0.6$,
with respect to the full population of $i'<22.5$ galaxies. This effect
on the radial selection of the survey, which we refer to as the Colour
Sampling Rate (\CSR) in the following, is only present below
$z=0.6$. Above this redshift, the colour selection has no impact on
the redshift selection and the sample becomes purely magnitude-limited
at $i'<22.5$ \citep{guzzo13}. If we weight the raw redshift
distribution by the global survey completeness function described in
the next sections, one obtains the $N(z)$ represented by the empty
histogram in Fig. \ref{fig1}. For convenience, we scaled down the
corrected $N(z)$ by $40\%$, the average effective survey sampling
rate, to aid the comparison between the shapes of the two
distributions. The difference in shape between these two $N(z)$ shows
the effect of incompleteness in the survey, which is only significant
at about $z>0.9$ \citep[see also][]{davidzon13}.

The observed redshift distribution in the sample can be well described
by a function of the form
\begin{equation}
N(z)=A \left(\frac{z}{z_0}\right)^\alpha
\exp\left(-\left(\frac{z}{z_0}\right)^\beta\right){CSR}(z), \label{eq:nz}
\end{equation} 
in units of $\rm{deg}^{-2} \cdot (\Delta z=0.03)^{-1}$ and where $(A,
z_0, \alpha, \beta)=(3.103, 0.191, 8.603, 1.448)$. The \CSR is the
incompleteness introduced by the VIPERS colour selection. It is
primarily a function of redshift and can be estimated from the ratio
between the number of galaxies with $i'<22.5$ satisfying the VIPERS
colour selection and the total number of galaxies with $i'<22.5$ as a
function of redshift. We calibrated this function using the VLT-VIMOS
Deep Survey Wide spectroscopic sample \citep[VVDS-Wide,][]{garilli08}
which has a CFHTLS-based photometric coverage and depth that is
similar to that of VIPERS, but which is free from any colour selection
\citep[see][for details]{guzzo13}. The \CSR is well described
  by a function of the form
\begin{equation}
{CSR}(z)=\left[\frac{1}{2}-\frac{{\rm
      erf}\left(b(z_t-z)\right)}{2}\right], \label{eq:csr}
\end{equation}
with $(b,z_t)=(17.465,0.424)$.

The fitting of $N(z)$ is important in measuring galaxy clustering: the
form of the mean redshift distribution must be followed accurately,
but features from large-scale structure must not be allowed to bias
the result.  We discuss this issue in detail in Section
\ref{sec:test}.

\section{Angular completeness}

\subsection{Slit assignment and footprint}

\begin{figure*}
\centering
\includegraphics[width=12cm]{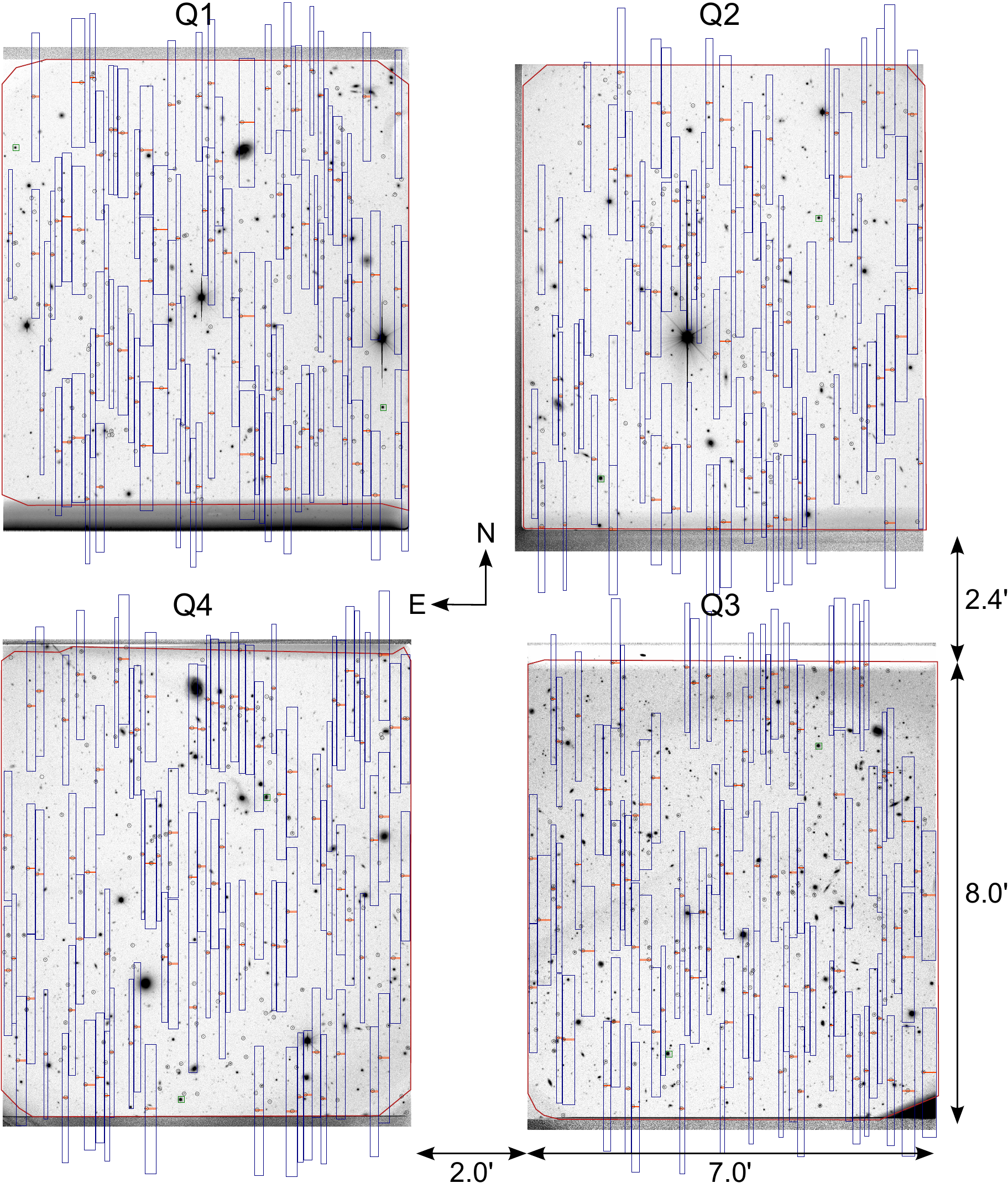}
\caption{Illustration of the slit assignment in pointing W1P082. The
  slits are shown in red and associated rectangles represent the
  typical dispersion of the spectra. All objects meeting the survey
  selection criteria (potential spectroscopic targets) are represented
  by black circles.}
\label{fig2}
\end{figure*}

To obtain a sample of several square degrees with VIMOS, one needs to
perform a series of individual observations or pointings. The VIPERS
strategy consists in covering the survey area with only one pass. This
has been done in order to maximise the volume probed. The survey
strategy and the fact that the VIMOS field-of-view is composed of four
quadrants delimited by an empty cross, create a particular footprint
on the sky which is reproduced in Figs \ref{fig3} and \ref{fig4}. In
each pointing, slits are assigned to a number of potential targets
which meet the survey selection criteria. This is shown in
Fig. \ref{fig2}, which illustrates how the slits are positioned in the
pointing W1P082. Given the surface density of the targeted population,
the multiplex capability of VIMOS, and the survey strategy, a fraction
of about $45\%$ of the parent photometric sample can be assigned to
slits. We define the fraction of target which have a measured spectrum
as the Target Sampling Rate (\TSR) and the fraction of observed
spectra with reliable redshift measurement as the Spectroscopic
Sampling Rate (\SSR). The number of slits assigned per pointing is
maximized by the SSPOC algorithm \citep{bottini05}, but the elongated size
of the spectra means that the resulting sampling rate is not uniform
inside the quadrants. The dispersion direction of the spectra in
VIPERS are aligned with the Dec direction and consequently, the
density of spectra along this direction is lower with respect to that
along the RA direction. This particular sampling introduces an
observed anisotropic distribution of pair separation, which has to be
accounted for in order to measure galaxy clustering correctly.

The two empty stripes between the four quadrants in each pointing
introduce a particular pattern in the measured correlation functions
if not accounted for. We correct for that by applying detailed binary
masks of the spectroscopic observations to a random sample of
unclustered objects, so that both data and random catalogues contain
no objects in these stripes. These masks account for the detailed
VIMOS field-of-view geometry as well as for the presence of vignetted
areas at the boundaries of the pointings. On top of these
spectroscopic masks, we apply a set of photometric masks which discard
areas where the parent photometry is affected by defects such as large
stellar haloes and where the survey selection is compromised
\citep[see][]{guzzo13}.

\subsection{Small-scale incompleteness}

\begin{figure}
\resizebox{\hsize}{!}{\includegraphics{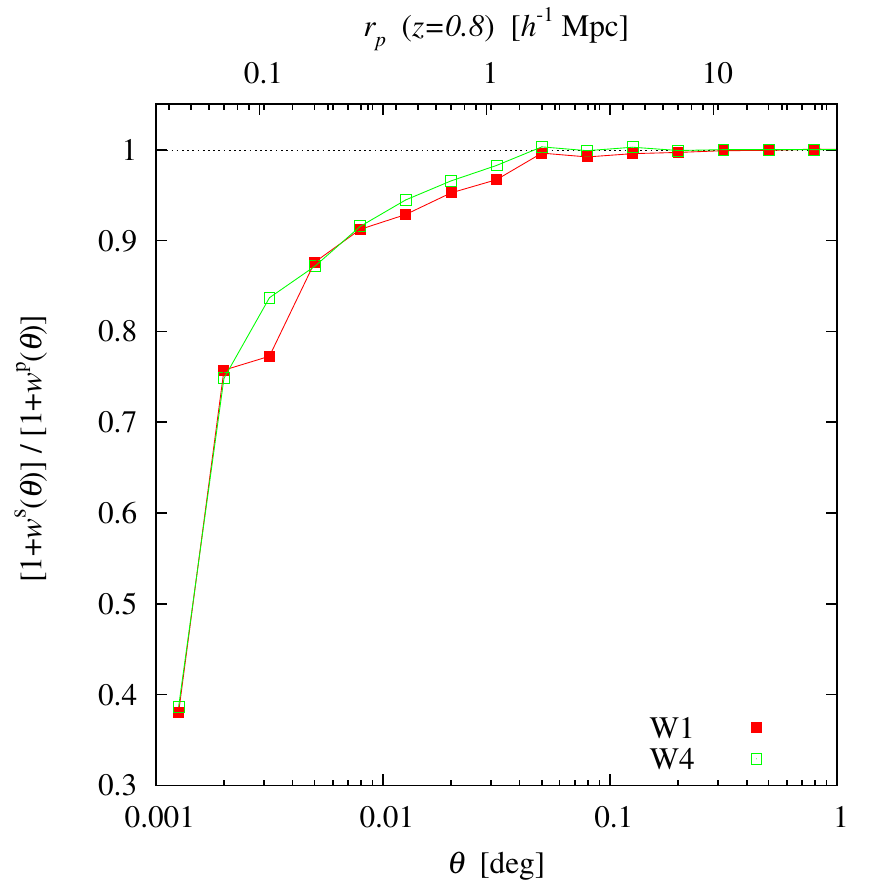}}
\caption{Completeness fraction of angular galaxy pairs due to the
  slit-spectroscopy strategy in the W1 and W4 fields for all galaxies
  at $0.5<z<1.0$. This has been obtained from the parent and
  spectroscopic sample angular correlation function.}
\label{fig5}
\end{figure}

We can characterise the amount of missing small-scale angular pairs
induced by the VIPERS spectroscopic strategy, by measuring the angular
pair completeness as a function of angular separation. This quantity,
defined as the ratio between the number of pairs in the spectroscopic
sample and that in the parent photometric sample, can be written in
terms of angular two-point correlation functions as \citep{hawkins03}
\begin{equation}
\frac{1}{w^A(\theta)}=\frac{1+w_s(\theta)}{1+w_p(\theta)}, \label{eq:angcomp}
\end{equation}
where $w_s(\theta)$ and $w_p(\theta)$ are respectively the angular
correlation function of the spectroscopic and parent samples. This
function is shown in Fig. \ref{fig5}. No significant difference is
seen between the W1 and W4 fields, as expected. The amount of missing
angular pairs is only significant below $\theta=0.03$ deg, which
corresponds to a transverse comoving scale of about $1\mhmpc$ at
$z=0.8$.

This fraction varies with redshift, although in practice we cannot
measure it at different redshifts since we do not have a measured
redshift for all galaxies in the parent sample. For this reason we use
the global $w^A(\theta)$ (averaged over all observed redshifts) to
correct for the small-scale angular incompleteness effect. We will
show in Section \ref{sec:test} that the level of systematic error
introduced by using $w^A(\theta)$ instead of $w^A(\theta|z)$ is very
small, of the order of a few percent. When measuring the angular
correlation functions, we include the completeness weights introduced
in the following section, in a similar way as for the three-dimensional
correlation function estimation.

It is important to mention that the small-scale angular incompleteness
effect is a general issue for large galaxy redshift surveys, in which
one has to deal with the mechanical constraints of multi-object
spectrographs and survey strategy. The incompleteness due to slit
assignment in VIPERS is to some extent similar to the fibre collision
problem in surveys using fibre spectroscopy such as 2dFGRS or SDSS,
while the magnitude of the effect is much more severe in our
case. Recently, a new method has been developed to accurately correct
for fibre collision \citep{guo12}. Although this method is quite
general, it is not applicable here. The exclusion between
spectroscopically observed objects in VIPERS is essentially
uni-directional, meaning that not all close pairs are
excluded. Therefore calculations such as that shown in Fig. \ref{fig3}
are possible from the set of one-pass observations, whereas the
correction scheme of \citep{guo12} can only be used for SDSS where
overlapping observations are included. Thus we need to revise the
correction methods developed for such surveys to apply them to VIPERS.

\subsection{Large-scale incompleteneness} \label{sec:esr}

\begin{figure*}
\centering
\includegraphics[width=17cm]{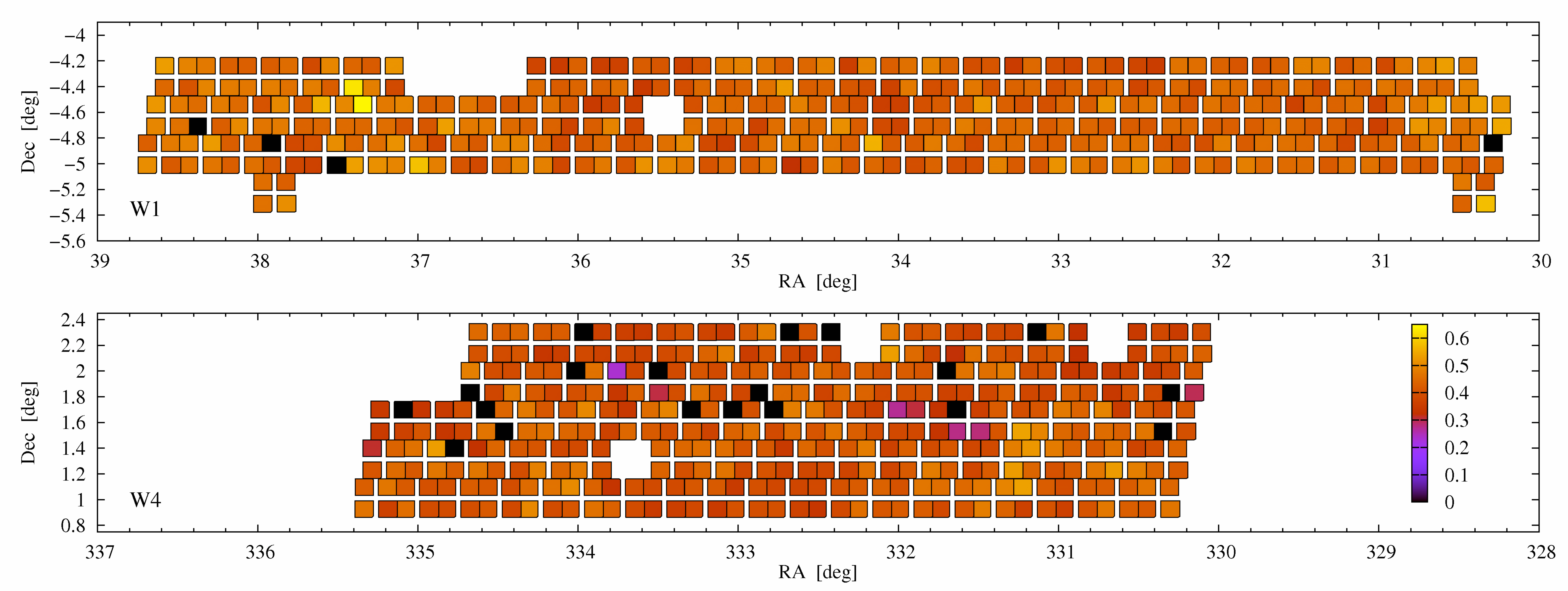}
\caption{Variations of the target success rate (\TSR) with
  quadrants. The \TSR quantifies our ability of obtaining spectra from
  the potential targets meeting the survey selection in the parent
  photometric sample. The quadrants filled in black correspond to
  failed observations where no spectroscopy has been taken.}
\label{fig3}
\end{figure*} 

\begin{figure*}
\centering
\includegraphics[width=17cm]{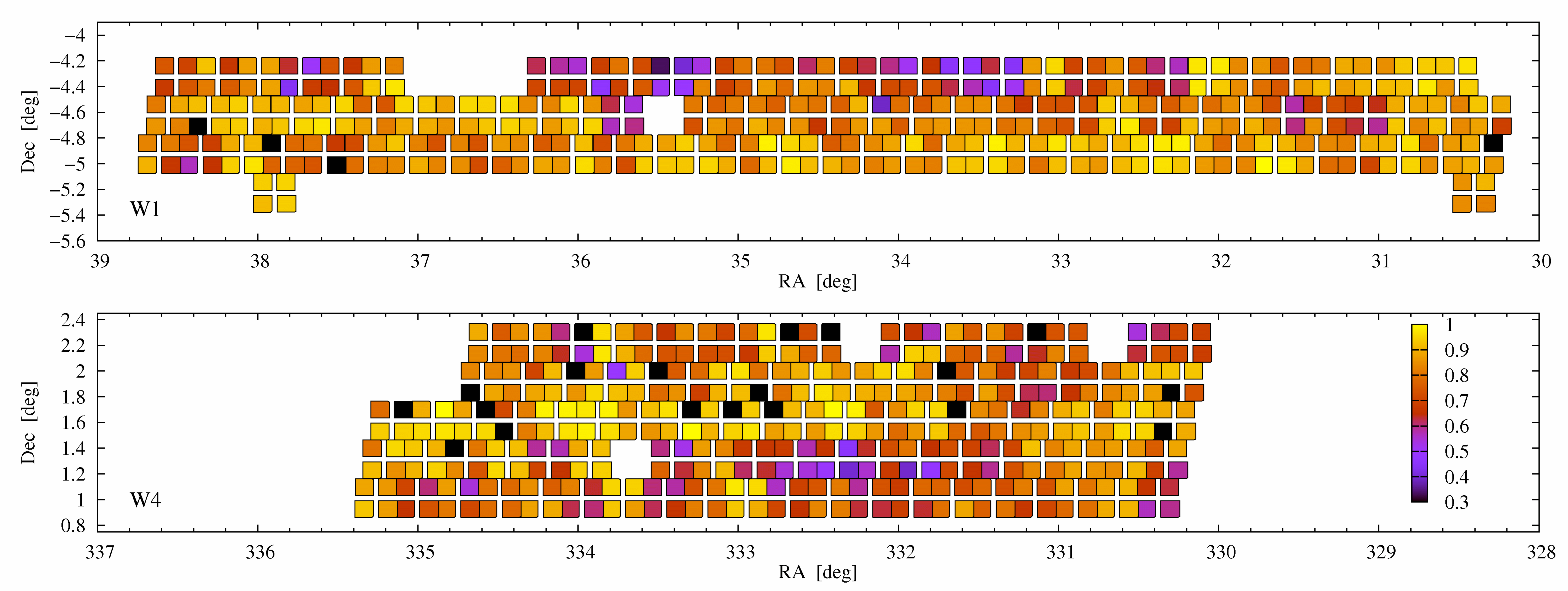}
\caption{Variations of the spectroscopic success rate (\SSR) with
  quadrants. The \SSR quantifies our ability of determining galaxy
  redshifts from observed spectra. The quadrants filled in black
  correspond to failed observations where no spectroscopy has been
  taken.}
\label{fig4}
\end{figure*} 

In addition to the non-uniform sampling inside the pointings, the
survey has variations of completeness from quadrant to quadrant. This
incompleteness is the combined effect of the \TSR and \SSR. The
latter, which characterises our ability of determining a redshift from
a galaxy spectrum, is determined empirically as the ratio between the
number of reliable redshifts and the total number of observed
spectra. The \TSR and \SSR in each quadrant are shown in in Figs
\ref{fig3} and \ref{fig4}. From these figures one can see clearly that
both \TSR and \SSR functions vary according to the position on the
sky, although the \SSR tends to have stronger variations. The
variations of \TSR reflect the changes in angular galaxy density in
the parent catalogue. Indeed, because of the finite maximum number of
slits that can be assigned and the fact that each quadrant has a
different number of potential targets, the less dense quadrants tend
to be better sampled than the denser ones. On the other hand,
variations in observational conditions from pointing to pointing
induce changes in \SSR. These different observational conditions
translate into variations of the signal-to-noise of the measured
spectra and so in our ability of extracting a redshift measurement
from them. These effects are taken into account in the clustering
estimation by weighting each galaxy according to the reciprocal of the
\TSR and \SSR.

\section{Clustering estimation} \label{sec:estimation}

We characterise the galaxy clustering in the VIPERS sample by
measuring the two-point statistics of the spatial distribution of
galaxies in configuration space. We estimate the two-point correlation
function $\xi(r)$ using the \citet{landy93} estimator
\begin{equation}
\xi(r)=\frac{GG(r)-2GR(r)+RR(r)}{RR(r)}, \label{eq:xir}
\end{equation}
where $GG(r)$, $GR(r)$, and $RR(r)$ are respectively the normalized
galaxy-galaxy, galaxy-random, and random-random number of pairs with
separation inside $[r-\Delta r/2,r+\Delta r/2]$. Note that here $r$ is
a general three-dimensional galaxy separation, not specifically the
real-space separation. This estimator minimises the estimation
variance and circumvent discreteness and finite volume effects
\citep{landy93,hamilton93}. A random catalogue must be constructed in
this estimator, whose aim is to accurately estimate the number density
of objects in the sample. It must be an unclustered population of
objects with the same radial and angular selection functions as the
data. In this analysis, we use random samples with 20 times more
objects than in the data to minimise the shot noise contribution in
the estimated correlation functions.

VIPERS has a complex angular selection function which has to be taken
into account carefully when estimating the correlation function. For
this, we weight each galaxy by the survey completeness weight, as well
as each pair by the angular pair weights described in the previous
section (Eq. \ref{eq:angcomp}). The survey completeness weights
correspond to the inverse of the effective sampling rate \ESR in each
quadrant $Q$, defined as
\begin{equation}
w(Q)=ESR^{-1}(Q)=(SSR(Q)\times TSR(Q))^{-1}. \label{eq:weight}
\end{equation}
By applying these weights we effectively up-weight galaxies in the
pair counts. It is important to note that here we keep the spatial
distribution of the random objects uniform across the survey
volume. We recall that survey completeness weights account for the
quadrant-to-quadrant variations of the survey completeness described
in Section \ref{sec:esr} but do not correct for the internal quadrant
incompleteness. For that we use the angular pair weights $w^A(\theta)$
which are applied to the GG pair counts. In principle the \ESR is also
a function of redshift and galaxy type
\citep[see][]{davidzon13}. However, given the statistics of the sample
it is impossible to measure the additional dependence of this function
on redshift and galaxy properties. Therefore, we decided to only
account for its quadrant-to-quadrant variations. We discuss the
accuracy of this approximation in Section \ref{sec:test}.

Additional biases can arise if the radial selection function exhibits
strong variations with redshift. The effect is particularly
significant for magnitude-limited catalogues covering a large range of
redshifts and in which the radial selection function rapidly drops at
high redshift. In that case, the pair counts is dominated by nearby,
more numerous objects: distant objects, although probing larger
volumes, will have less weight. To account for this we use the minimum
variance estimator of \citet{davis82} for which the galaxy counts are
essentially weighted by the inverse of the volume probed by each
galaxy. This weighting scheme, usually referred as the $J_3$
weighting, is defined as \citep{hamilton93}
\begin{equation}
w^{J_3}(z,s)=\frac{1}{1+\bar{n}(z) 4\pi J_3(s)}\, ,
\end{equation}
where $z$ is the redshift of the object, $s$ is the redshift-space
pair separation, $\bar{n}(r)$ the galaxy number density at $z$ and
$J_3(s)$ is defined as
 \begin{equation}
J_3(s)=\int_0^s s^{\prime 2} \xi(s^\prime)ds^\prime.
\end{equation}
Each pair is then weighted by,
\begin{equation}
w^{J_3}_{ij}=w^{J_3}_i(z_i,s_{ij})w^{J_3}_j(z_j,s_{ij}).
\end{equation}
However, we find that applying $J_3$ weighting does not significantly
change the amplitude and shape of the correlation function in our
sample, and tends to produce noisy correlation functions especially for
high-redshift sub-samples. We thus decided not to apply this
correction in this analysis.

The final weight assigned to $GG$, $GR$, and $RR$ pairs combine the
survey completeness and angular pair weights as
\begin{align}
GG(r)&=\sum_{i=1}^{N_G}\sum_{j=i+1}^{N_G}w_i(Q_i)w_j(Q_j)w^A(\theta_{ij})\Theta_{ij}\left(r\right) \\
GR(r)&=\sum_{i=1}^{N_G}\sum_{j=1}^{N_R}w_i(Q_i)\Theta_{ij}\left(r\right) \\
RR(r)&=\sum_{i=1}^{N_R}\sum_{j=i+1}^{N_R}\Theta_{ij}\left(r\right) \, ,
\end{align}
where $\Theta_{ij}(r)$ is equal to unity for $r_{ij}$ in $[r-\Delta
  r/2,r+\Delta r/2]$ and null otherwise. 

We measure correlation functions using both linear and logarithmic
binning. We define the separation associated with each bin as the bin
centre and as the mean pair separation inside the bin, respectively
for the linear and logarithmic binning \citep{zehavi11}. The latter
definition is more accurate than using the bin centre, in particular at
large $r$ when the bin size is large.

The galaxy real-space correlation function $\xi(r)$ is not directly
measurable from redshift survey catalogues because of galaxy peculiar
velocities that affect redshift measurements. Galaxy peculiar
velocities introduce distortions in the galaxy clustering pattern and
as a consequence we can only measure redshift-space quantities. We
measure the anisotropic redshift-space correlation function
$\xi(r_p,\pi)$ in which the redshift-space galaxy separation vector
has been divided in two components, $r_p$ and $\pi$, respectively
perpendicular and parallel to the line-of-sight \citep{fisher94}. This
decomposition, which assumes the plane-parallel approximation, allows
us to isolate the effect of peculiar velocities as these modify only
the component parallel to the line-of-sight. Redshift-space
distortions can then be mitigated by integrating $\xi(r_p,\pi)$ over
$\pi$, thus defining the projected correlation function
\begin{equation}
w_p(r_p)=\int^{\pi_{\rm max}}_{-\pi_{\rm max}} \xi(r_p,\pi)d\pi.
\end{equation}
We measure $w_p(r_p)$ using an optimal value of $\pi_{\rm
  max}=40\mhmpc$, allowing us to reduce the underestimation of the
amplitude of $w_p(r_p)$ on large scales and at the same time to avoid
including noise from uncorrelated pairs with separations of
$\pi>40\mhmpc$. The projected correlation function allows us to
measure real-space clustering (but see the later parts of Section
\ref{sec:test}). To combine the correlation function measurements from
the two fields, we measure the mean of one plus the correlation
functions in W1 and W4 weighted by the square of the number density,
so that the combined correlation function $\xi(r_p,\pi)$ is obtained
from
\begin{equation}
1+\xi(r_p,\pi)=\frac{n^2_{W1}(1+\xi_{W1}(r_p,\pi))+n^2_{W4}(1+\xi_{W4}(r_p,\pi))}{n^2_{W1}+n^2_{W4}},
\end{equation}
where $n_{W1}$ and $n_{W4}$ are the observed galaxy number densities
in the W1 and W4 fields respectively.

\section{Tests of the clustering estimation} \label{sec:test}

\subsection{Simulation data} \label{sec:mocks}

To test the robustness of our clustering estimation we make use of a
large number of mock galaxy samples, which are designed to be a
realistic match to the VIPERS sample. We create two sets of mock
samples based on the Halo Occupation Distribution (HOD)
technique. These two sets only differ by the input halo catalogue that
has been used. In the first set of mocks, we used the haloes extracted
from the MultiDark dark matter N-body simulation \citep{prada12}. This
simulation, which assumes a flat $\Lambda {\rm CDM}$ cosmology with
$(\Omega_m,~\Omega_\Lambda,~\Omega_b,~h,~n,~\sigma_8) =
(0.27,~0.73,~0.0469, ~0.7,~0.95,~0.82)$, covers a volume of $1\mhgpcc$
using $N=2048^3$ particles.  In the simulation, the haloes have been
identified using a friends-of-friends algorithm with a relative
linking length of $b=0.17$ times the inter-particle separation
(i.e. $0.083\mhmpc$) . The mass limit to which halo catalogues are
complete is $10^{11.5}\mhmsun$. Because this limiting mass is too
large to host the faintest galaxies observed with VIPERS, we use the
method of \citet{delatorre13} to reconstruct haloes below the
resolution limit. This method is based on stochastically resampling
the halo number density field using constraints from the
conditional halo mass function. For this, one needs to assume the
shapes of the halo bias factor and halo mass function at masses below
the resolution limit and use the analytical formulae obtained by
\citet{tinker08,tinker10}. With this method we are able to populate
the simulation with low-mass haloes with a sufficient accuracy to have
unbiased galaxy two-point statistics in the simulated catalogues
\citep[see][for details]{delatorre13}. The minimum reconstructed halo
mass we consider for the purpose of creating VIPERS mocks is
$10^{10}\mhmsun$.

We then apply to the complete halo catalogues the algorithm presented
in \citet{carlson10} to remap halo positions and velocities in the
initial simulation cube onto a cuboid of the same volume but different
geometry. This is done to accommodate a maximum number of disjoint
VIPERS W1 and W4 fields within the $1\mhgpcc$ volume of the
simulation.  This process allows us to create 26 and 31 independent
lightcones for W1 and W4 respectively over the redshift range
$0.4<z<1.3$. The lightcones are built by considering haloes from the
different snapshots, disposing them according to their distance from
the coordinate origin of the lightcone. The lightcones are then
populated with galaxies using the HOD technique. In this process, we
populate each halo with galaxies according to its mass, the mean
number of galaxies in a halo of a given mass being given by the
HOD. It is common usage to differentiate between central and satellite
galaxies in haloes. While the former are put at rest at halo centres,
the latter are randomly distributed within each halo according to a
NFW radial profile. The halo occupation function and its dependence on
redshift and luminosity/stellar mass must be precisely chosen in order
to obtain mock catalogues with realistic galaxy clustering
properties. We calibrated the halo occupation function directly on the
VIPERS data. We performed an analytic HOD modelling of the projected
correlation function for different samples selected in luminosity and
redshift that we will present in Section \ref{sec:realclus}. We obtain
from this a series of HOD parameters at different redshifts and for
different cuts in $B$-band absolute magnitude, which we then
interpolate to obtain a general redshift- and $B$-band absolute
magnitude-dependent halo occupation function $\langle N_{\rm
  gal}(m|z,M_B)\rangle$. We use the latter function to populate the
haloes with galaxies. Finally, we add velocities to the galaxies and
measure their redshift-space positions. While the central galaxies are
assigned the velocity of their host halo, satellite galaxies have an
additional random component for which each Cartesian velocity
component is drawn from a Gaussian distribution with a standard
deviation that depends on the mass of the host halo. Details about the
galaxy mock catalogue construction are given in Appendix A.

The second set of mocks that we built is based on halo catalogues
created with the Pinocchio code\footnote{We have used in this analyis
  a new version of this code, optimised to work on massively parallel
  computers, which is described in \citet{monaco13}.}
\citep{monaco02}. This code follows the evolution of a set of
particles on a regular grid using an ellipsoidal model to compute
collapse times and identify dark matter halos, and the Zel'dovich
approximation to displace the haloes from their initial
position. While the recovery of haloes works well on an
object-by-object basis, their positions and velocities on scales below
$10\mhmpc$ suffer by the lack of accuracy of the Zel'dovich
approximation. The halo positions and velocities obtained with this
method are less accurate than those from the N-body simulation, and
the halo clustering is generally underestimated on scales below
$3\mhmpc$ \citep[e.g.][]{monaco02}. However this approach has the
advantage of being very fast and can be used to generate a large
number of independent halo catalogue realizations. We created $200$
independent halo mock realizations assuming the same cosmology as the
MultiDark N-body simulation. The remaining steps in generating galaxy
mock samples are similar to those used for the mocks based on the
MultiDark simulation. The only difference is that here we do not need
to divide each simulation into sub-volumes to generate different
lightcones: we can directly create volumes of the size of the
lightcones.

The final step in obtaining fully realistic VIPERS mocks is to add the
detailed survey selection function. The procedure that we follow is
similar to that used in the VVDS and zCOSMOS surveys, which were also
based on VIMOS observations
\citep[][]{meneux06,iovino10,delatorre11}. We start by applying the
magnitude cut $i'<22.5$ and the effect of the colour selection on the
radial distribution of the mocks. The latter is done by depleting the
mocks at $z<0.6$ so as to reproduce the \CSR. The mock catalogues that
we obtain are then similar to the parent photometric sample in the
data. We next apply the slit-positioning algorithm with the same
setting as for the data. This allows us to reproduce the VIPERS
footprint on the sky, the small-scale angular incompleteness and the
variation of \TSR across the fields. Finally, we deplete each quadrant
to reproduce the effect of the \SSR. Thus we are able to produce
realistic mock galaxy catalogues that contain the detailed survey
completeness function and observational biases of VIPERS.

\subsection{Effects of systematics on the correlation function}

\subsubsection{Effects related to the radial selection function}

We first study the impact on our correlation function measurements of
using different methods to estimate the radial selection. A key aspect
in three-dimensional clustering estimation is to have a smooth and
unbiased redshift distribution from which the random sample can be
drawn. In particular, when the data sample used to estimate the radial
distribution is not very large, one generally has to deal with strong
features associated with prominent structures; these must not be allowed
to induce spurious clustering in the random sample.

There are several empirical methods for avoiding this problem. One can
for instance interpolate the binned observed distribution using cubic
splines, filter the observed distribution with a kernel sufficiently
large to erase the strong features in the distribution, or fit the
observed distribution with a smooth template $N(z)$ and then randomly
sample it. In general most of the methods are parametric and have to
be calibrated. An alternative non-parametric method is the $V_{\rm
  max}$ method. This method consists in randomly sampling the maximum
volumes $V_{\rm max}$ probed by each galaxy in the survey
\citep[e.g.][]{kovac10,cole11}. The $V_{\rm max}$ value for each
galaxy corresponds to the volume between the minimum and the maximum
redshifts $z_{\rm min}$ and $z_{\rm max}$ at which the galaxy is
observable in the survey.

\begin{figure}
\resizebox{\hsize}{!}{\includegraphics{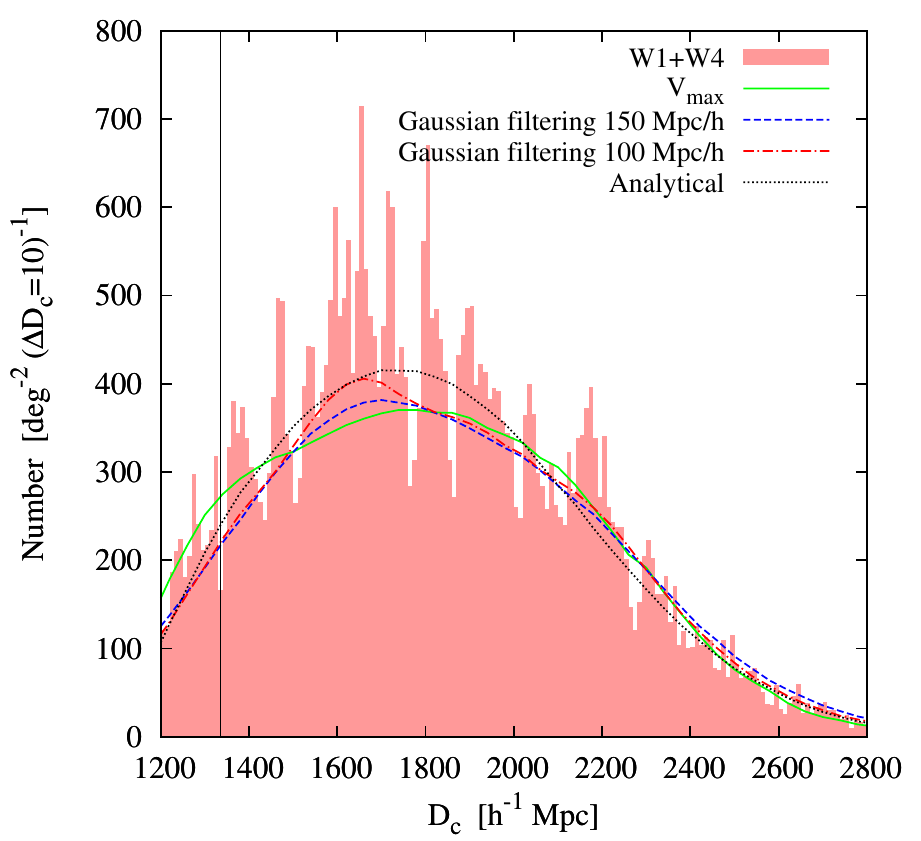}}
\caption{Comparison of different estimators of the radial distribution
  in the combined W1+W4 sample. The filled histogram shows the number
  of galaxies in fine bins of radial comoving distance. The different
  curves correspond to random radial distribution realizations
  normalized to the number of objects in the data, obtained using the
  $V_{\rm max}$ (solid), Gaussian filtering (dashed and dot-dashed),
  or analytical (dotted) methods. The vertical line shows the minimum
  redshift considered in this analysis, i.e. $z=0.5$.}
\label{fig6}
\end{figure}

\begin{figure}
\resizebox{\hsize}{!}{\includegraphics{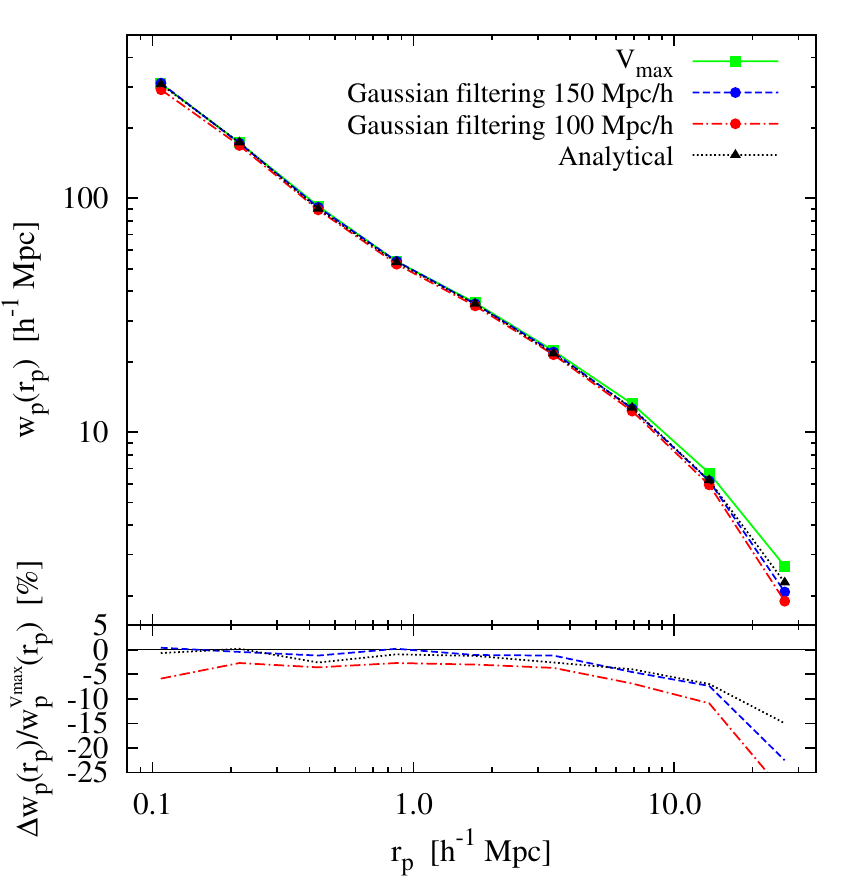}}
\caption{Impact of the use of different estimators of the radial
  distribution on the shape of the projected correlation function. The
  projected correlation functions obtained using the $V_{\rm max}$
  (solid), Gaussian filtering (dashed and dot-dashed), or analytical
  (dotted) methods are shown in the top panel, while the relative
  fractional differences with respect to the $V_{\rm max}$ method are
  presented in the bottom panel.}
\label{fig7}
\end{figure}

Fig. \ref{fig6} applies three such approaches to estimate the galaxy
radial distribution in the combined W1+W4 sample: the analytical
$N(z)$ of Eq. \ref{eq:nz}; the Gaussian filtering method; and the
$V_{\rm max}$ method. This figure shows the recovered $N(D_c)$ in the
random sample with each method, with $D_c$ being the radial comoving
distance; in practice we work with $N(D_c)$ instead of $N(z)$.  We
find that the methods give different estimates of the radial
distribution. In the case of the the Gaussian filtering, a kernel size
of $150\mhmpc$ is needed to smear out the peaks in the distribution,
otherwise the recovered $N(D_c)$ is still affected by large structures
in the field -- particularly by that at $D_c\simeq1600\mhmpc$. As
expected, the filtering method tends to artificially broaden the
$N(D_c)$ distribution, whereas the analytical and $V_{\rm max}$
methods are much smoother by construction and do not broaden the
$N(D_c)$. We find that the $V_{\rm max}$ estimate shows a slightly
flatter distribution at the level of the peak of the distribution,
which seems visually to be more consistent with the data. In
Fig. \ref{fig7} we show the effect of using these different estimates
of the radial distribution on the shape of the measured correlation
function.  Gaussian filtering with a kernel size of $150\mhmpc$ and
analytical $N(z)$ estimates both yield slightly smaller amplitudes of
the projected correlation function on scales of above $10\mhmpc$ than
the $V_{\rm max}$ method. Gaussian filtering with a kernel size of
$100\mhmpc$ globally underestimates the clustering amplitude on
$w_p(r_p)$ as expected, by about $5\%$. The analytical and $V_{\rm
  max}$ methods give very similar answers, except on scales above
$5\mhmpc$ where the former tends to produce a smaller clustering
amplitude by $5-15\%$ with respect to the latter. This comparison
shows that the $V_{\rm max}$ method is more robust as it uniquely
allows us to restore some correlation signal at large separation. For
this reason and the fact that it is non-parametric we finally decided
to use the $V_{\rm max}$ estimate to measure two-point correlation
functions.

\subsubsection{Effects related to the angular selection function}

The most crucial aspect of the galaxy clustering estimation in VIPERS
is to account for the angular selection function. We test our
methodology and the different assumptions discussed in Section
\ref{sec:estimation} using the MultiDark mock samples. We measure the
accuracy with which we can estimate the two-point correlation
function, by confronting the two-point correlation functions measured
in the parent catalogues with those measured in the observed mocks
when different completeness corrections are included. We measure the
average relative difference between the corrected observed mocks and
the parent measurement for different statistics. For this test, we
consider two galaxy samples encompassing respectively all galaxies in
the redshift intervals $0.5<z<0.75$ and $0.75<z<1.0$, using the same
redshift distribution in the parent and observed mock samples to
construct the radial selection function of the random sample.

\begin{figure}
\resizebox{\hsize}{!}{\includegraphics{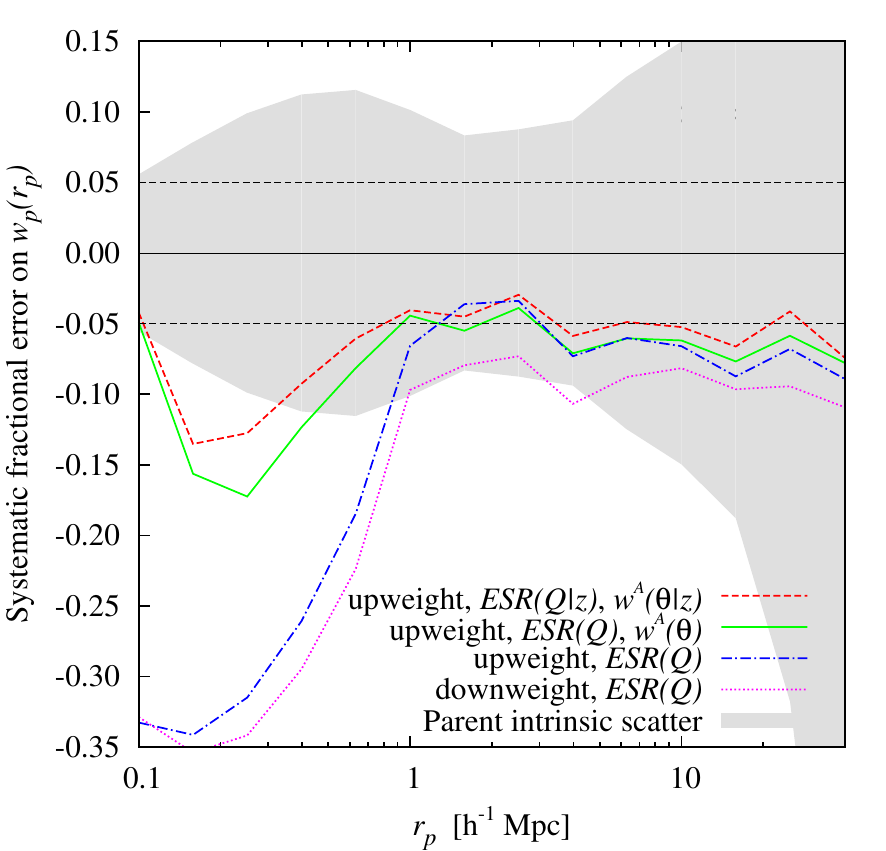}}
\vglue-0.8em
\resizebox{\hsize}{!}{\includegraphics{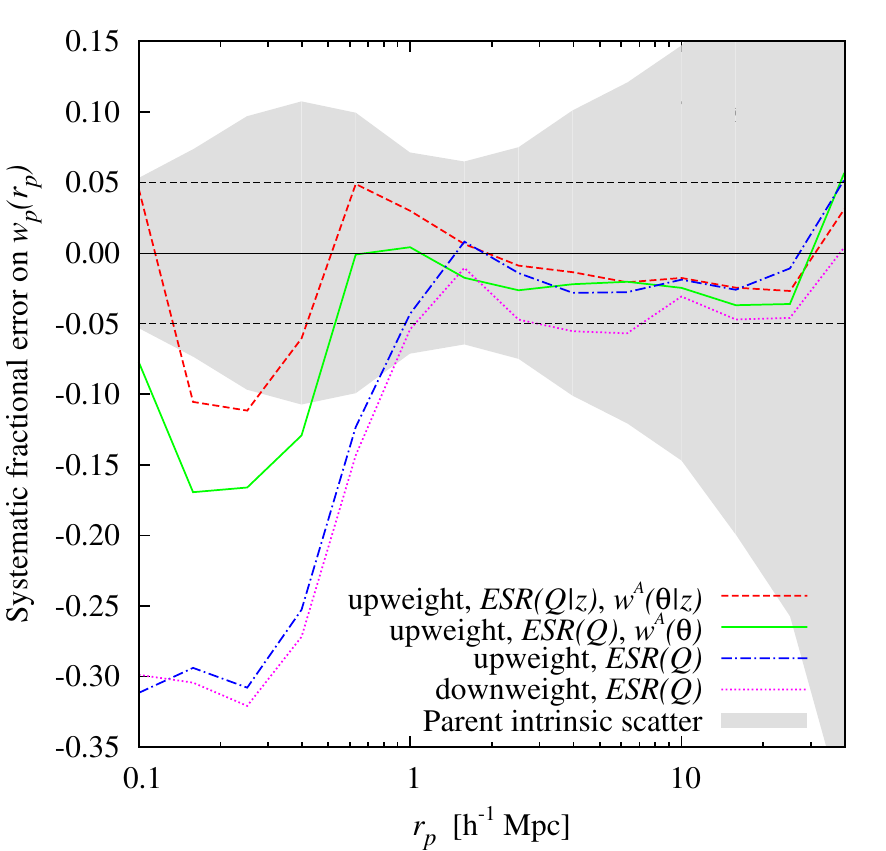}}
\caption{Systematic error on the projected correlation function and
  impact of different corrections. This is calculated considering all
  VIPERS galaxies in the redshift intervals $0.5<z<0.75$ (top panel)
  and $0.75<z<1$ (bottom panel).}
\label{fig8}
\end{figure}

It is common usage in clustering analysis to account for the angular
survey completeness by down-weighting the random pair counts. This is
usually done by keeping the galaxy counts unweighted and depleting the
random sample so as to reproduce the survey angular completeness. The
same effect can be achieved by using a uniform angular distribution of
random objects but weighting each of them by the inverse of the weight
defined in Eq. \ref{eq:weight}. If we do that and set all the angular
pair weights to unity, we obtain the systematic error on $w_p(r_p)$
shown with the dotted curves in Fig. \ref{fig8}. We concentrate first
on the results in the interval $0.5<z<0.75$. We can see in this figure
that the recovered clustering with this method is underestimated by
about $10\%$ at about $1<r_p<20\mhmpc$, and then drops rapidly to
$35\%$ below. The strong underestimation on small scales is due to the
small-scale angular incompleteness effect inside the quadrants. The
approach of modulating the random density is dubious in the context of
VIPERS, since it treats the sampling variations as a pattern imposed
on the large-scale structure. But because of the VIMOS slit
allocation, these variations are strongly coupled with the true
clustering (i.e. the observed sky distribution of VIPERS galaxies is
rather uniform). It is therefore safer if we keep the random sample
uniform but upweight the galaxies as described in section
\ref{sec:estimation}. In this case, we obtain the dot-dashed lines in
Fig. \ref{fig8}: these represent an improved estimation of $w_p(r_p)$,
reducing the underestimation by $5-6\%$. As expected, further
including the angular pairs weights permits us to remedy in part the
underestimation on scales below $1\mhmpc$, where the systematic error
reaches $15\%$ (solid lines).

So far, we have used the global survey completeness and angular
weights, i.e. neglecting the redshift dependence. As an exercise we
use the redshift information from the parent mocks to compute the true
redshift-dependent weights and we obtain the dashed lines in the
figure. Including the redshift dependence in the weights has the
effect of improving the recovery of the projected correlation function
by about $2\%$ over all probed scales. However, this improvement is
rather modest -- indicating that the use of the redshift-independent
weights is a good approximation. Our best estimate of $w_p(r_p)$
therefore allows us to recover the true correlation function of the
mocks at $0.5<z<0.75$ with about $7\%$ and $16\%$ underestimation
respectively above and below $1\mhmpc$. In the redshift interval
$0.75<z<1$ (shown in th bottom panel of Fig. \ref{fig8}), we find the
same behaviour except that the correlation function is globally better
recovered with an underestimation smaller than $2-3\%$ at
$r_p>0.6\mhmpc$ with the best method.

This test demonstrates that our methodology gives an accurate estimate
of the galaxy clustering in VIPERS, even if there remains some
residual systematic errors of up to $7\%$ on the scales above 1\hmpc
and $15\%$ on smaller scales. We find that the effect varies with
redshift, being more important at the lowest redshifts probed by
VIPERS. Overall these systematics remain within the Poisson plus
sample variance errors, shown with shaded regions in Fig. \ref{fig8}
and obtained from the standard deviation of \wprp among the parent
mock catalogues.

\subsection{Impact of possible residual zero-point uncertainties in the photometry}

At the time of writing, photometry from the latest CFHTLS release
(T0007) has become available \citep{hudelot12}. We have compared
magnitudes and colours of objects in the VIPERS sample with the new
CFHTLS-T0007 photometry. For VIPERS, the most important feature of
T0007 compared to previous releases is that each tile in the CFHTLS
has now been rescaled to an absolute calibration provided by a new
photometric pre-survey taken at CFHT for this purpose. In addition, in
order to ensure that seeing variations between tiles and filters are
correctly accounted for, this scaling has been done using aperture
fluxes that are rescaled based on the seeing on each individual tile;
detailed tests at Terapix have shown that mag\_auto magnitudes, which
are affected by seeing variations, are not sufficiently precise for
the percent-level photometric accuracy that is the objective of T0007.

An important consequence of this work for VIPERS is that the effect of
seeing variation and photometric calibration errors are now cleanly
separated; the stellar-locus fitting technique used to define the
VIPERS selection using colours based on mag\_auto magnitudes mixes
both these effects. To estimate the size of colour and magnitude
offsets between T0007 and the actual VIPERS selection (based on T0005)
colours of stars on each VIPERS tiles measured from Terapix IQ20
magnitudes (used to calibrate T0007) and from mag\_auto magnitudes in
both releases have been compared. We find that these offsets shift the
colour-colour locus we devised to remove lower-redshift $z<0.5$
galaxies \citep{guzzo13}.

\begin{figure}
\resizebox{\hsize}{!}{\includegraphics{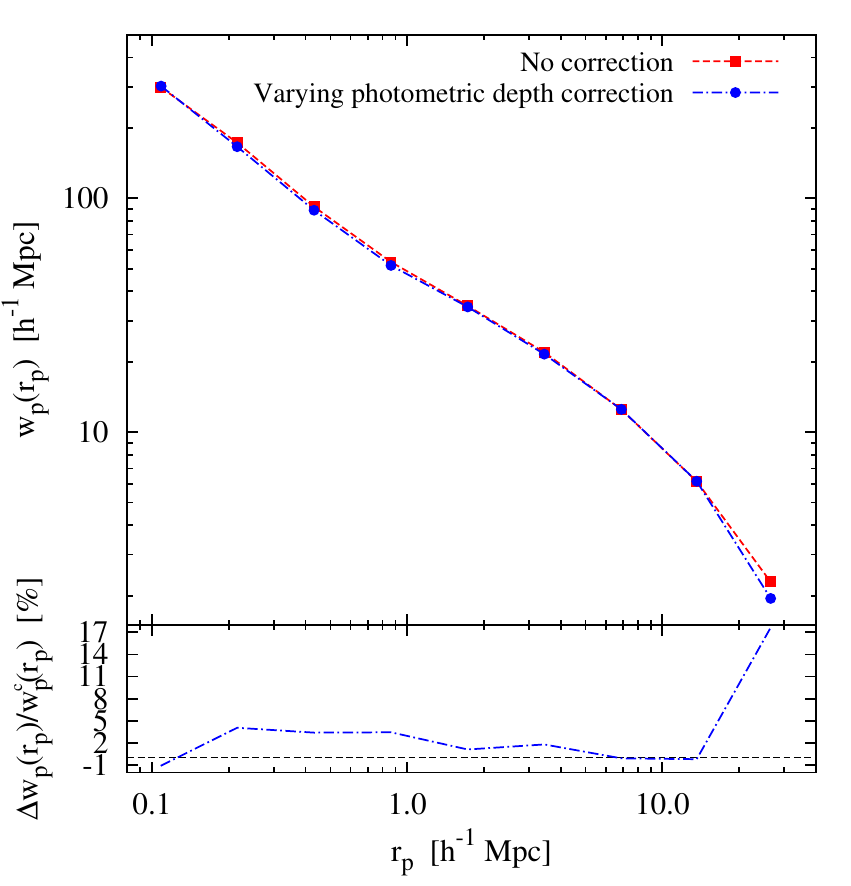}}
\caption{Effect of correcting for quadrant-to-quadrant variations of
  the colour selection in the estimation of the projected correlation
  function, when using the CFHTLS-T0007 sample as the reference
  photometric catalogue. The top panel shows the projected correlation
  functions with and without the correction applied, while the bottom
  panel presents their relative fractional difference.}
\label{fig10}
\end{figure}

We test the effect of these possible variations of the colour
selection across the fields in the context of galaxy clustering
estimation. For this we use photometric redshifts and quantify the
variations in $N(z)$ due to tile-to-tile variations of the colour
selection, assuming the T0007 photometry as the reference. When
comparing the $N(z_{phot})$ in the different tiles, we find that the
redshift distribution varies in shape and amplitude at $z<0.6$ but
only in amplitude above. The typical amplitude variations are of the
order of about $5\%$ \citep{guzzo13}. We then measure the ratio
between the $N(z)$ per tile and that averaged over the fields and use
it as a redshift-dependent correction factor. To test how these
variations of the colour selection affect the measured correlation
function, we vary the $N(z)$ in the random sample for each quadrant
using the correction factor previously defined on the averaged $N(z)$.
The projected correlations obtained with and without this correction
are shown in Fig. \ref{fig10}.

We can see that the correction has the effect of decreasing the
amplitude of the projected correlation function by about $2-4\%$ on
scales below $10\mhmpc$. We find a similar effect on the
redshift-space angle-averaged correlation function $\xi(s)$. The
amplitude and direction of the systematic effect follows our
expectations, since spurious tile-to-tile fluctuations, if not
properly corrected, enhance the amplitude of clustering. This test
suggests that indeed such tile-to-tile variations of colour selection
are present in the data. It is interesting to note that this
systematic effect goes in the opposite direction to the effects of
slit-positioning and associated incompleteness. In the end, because
this possible effect remains very small, we do not attempt to correct
it for the clustering analysis.

\section{Real-space clustering}\label{sec:realclus}

\begin{figure}
\resizebox{\hsize}{!}{\includegraphics{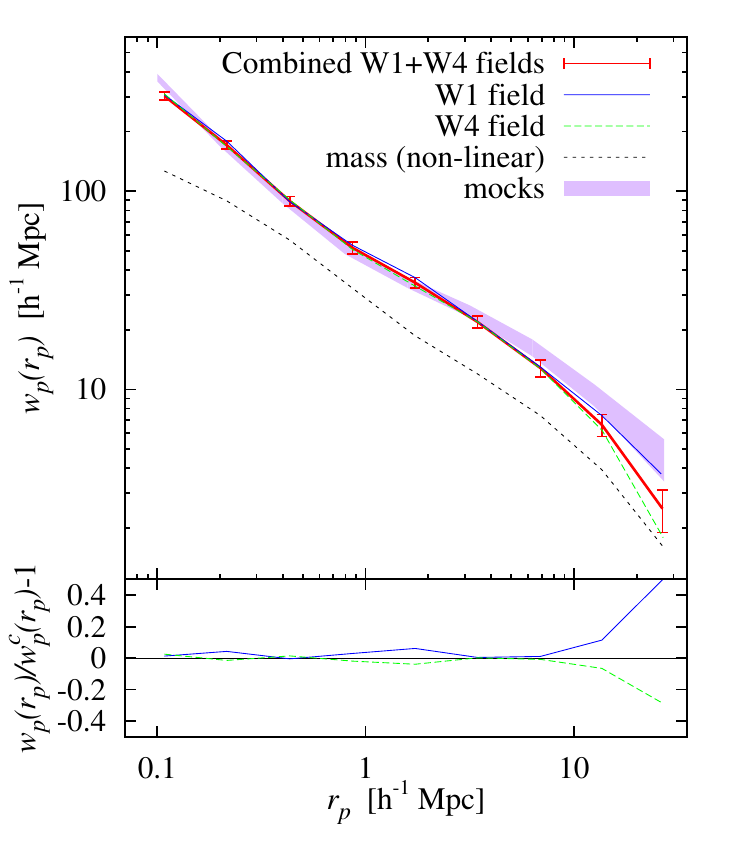}}
\caption{Top panel: projected correlation functions of VIPERS galaxies
  in the redshift interval $0.5<z<1$ for the individual W1 and W4
  fields as well as for the combined sample. As a comparison, the $\pm
  1\sigma$ dispersion among the mean \wprp in the mocks is shown with
  the shaded region and the non-linear mass prediction in the assumed
  cosmology with the dotted curve. Bottom panel: relative
    difference between the measured \wprp in the W1 and W4 fields and
    the combined projected correlation function $w^c_p(r_p)$.}
\label{fig11}
\end{figure}

Before studying redshift-space distortions in VIPERS, we begin by
looking at the clustering in real space.  The projected correlation
function for all galaxies in the redshift range $0.5<z<1$ is shown in
Fig. \ref{fig11}. It is measured in logarithmic bins of $\Delta \log
r_p=0.2$ over the scales $0.1<r_p<30\mhmpc$. The error bars are
estimated from the MultiDark mocks.

The measured \wprp functions in the W1 and W4 fields are very similar,
in particular on scales below 5\hmpc.  The combined projected
correlation function in this redshift interval gives an accurate probe
of the clustering up to scales of about 30\hmpc. We can compare the
galaxy projected correlation function to predictions for the mass
non-linear correlation function and thus estimate the global effective
linear bias of these galaxies. We use the HALOFIT \citep{smith03}
prescription for the non-linear mass power spectrum to compute the
projected correlation function of mass at the mean redshift of the
sample. By comparing the amplitudes of the measured galaxy and
predicted mass correlations on scales of $r_p>1.7\mhmpc$
($r_p>3.4\mhmpc$), and assuming a linear biasing relation of the form
$\smash{w_p^{\rm gal}=b^2_L w_p^{\rm mass}}$, we obtain a linear bias
of $b_L=1.35\pm0.02$ ($b_L=1.33\pm0.02$).

\begin{figure}
\resizebox{\hsize}{!}{\includegraphics{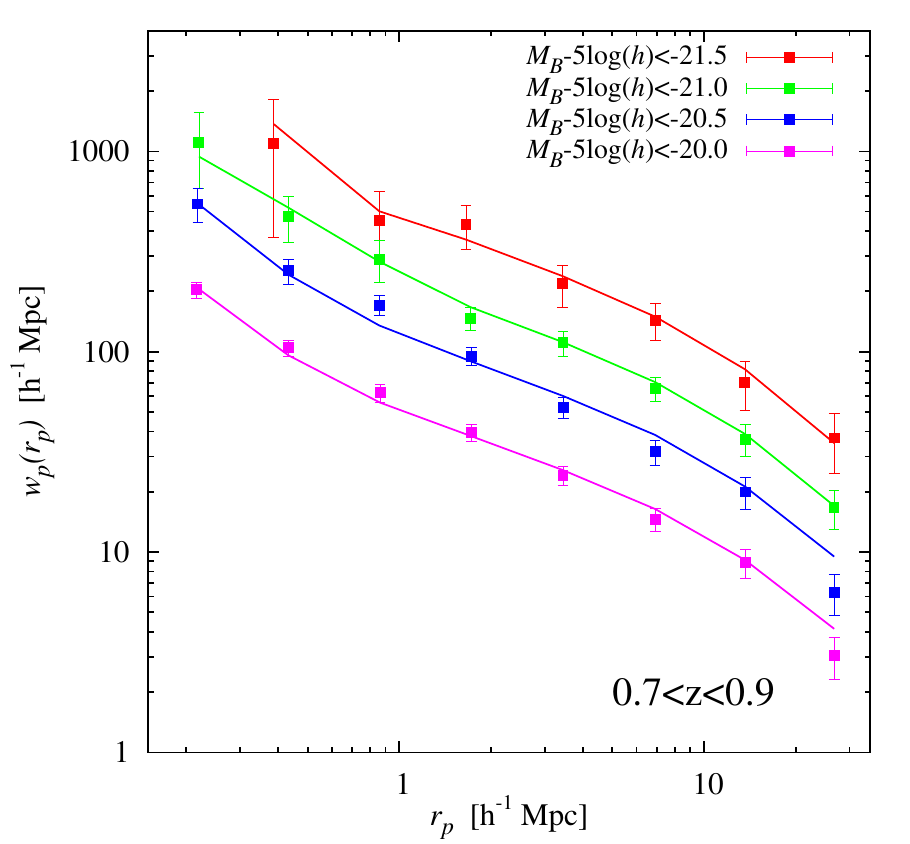}}
\caption{Measured and best-fitting HOD model projected correlation
  functions \wprp, for different luminosity-threshold subsamples of
  galaxies at $0.7<z<0.9$. The \wprp for the $M_B-5\log(h)<-20.5$,
  $M_B-5\log(h)<-21.0$, and $M_B-5\log(h)<-21.5$ cases have been
  multiplied by respectively $2$, $3$, and $4$ to improve the clarity
  of the figure.}
\label{fig12}
\end{figure}

\begin{figure}
\resizebox{\hsize}{!}{\includegraphics{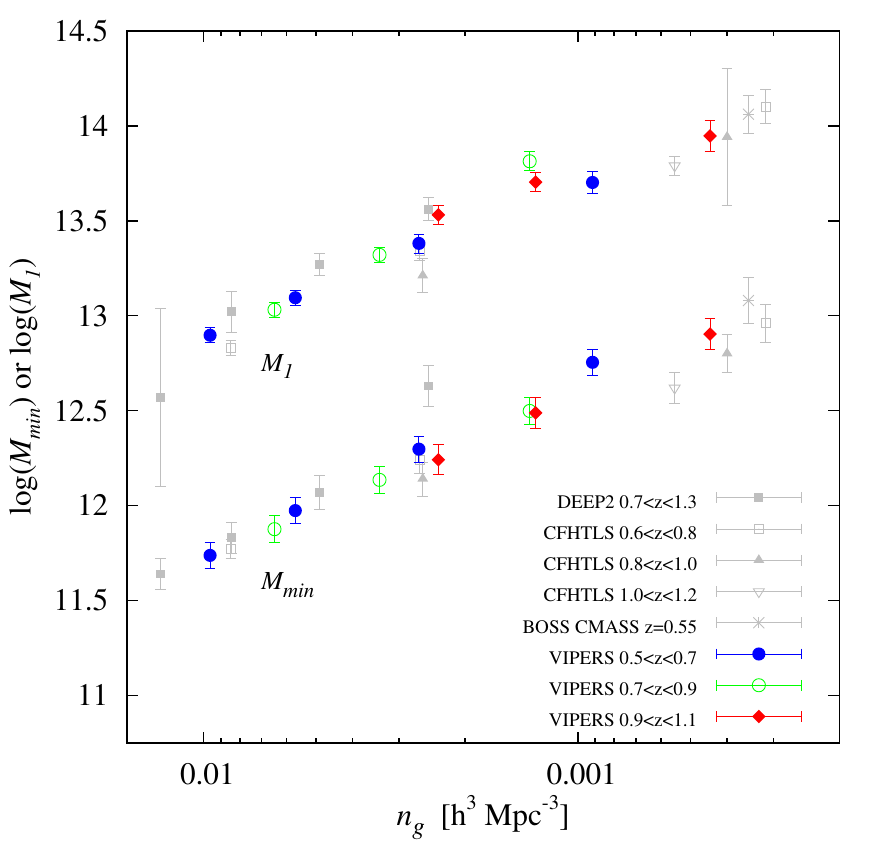}}
\caption{The dependence of $M_{\rm min}$ and $M_{1}$ HOD parameters on
  redshift and absolute magnitude threshold in $B$-band. The absolute
  magnitude threshold samples are plot in terms of their implied
  number density $n_g$. The VIPERS results are compared to the
  pervious measurements performed in the DEEP2 \citep{zheng07}, BOSS
  \citep{white11}, and CFHTLS \citep{coupon12} surveys.}
\label{fig13}
\end{figure}
                                                                      
In order to make a detailed interpretation of the observed clustering
of galaxies and produce realistic mock samples of the survey, we model
our \wprp measurements within the context of the HOD
\citep{seljak00,peacock00,berlind02,cooray02}. This method defines the
mean distribution of galaxies within haloes; under the assumption of
the abundance, large-scale bias, and density profile of haloes, one
can then completely specify the clustering of galaxies and predict
\wprp. We define four $B$-band absolute magnitude-threshold samples in
the redshift bin $0.7<z<0.9$ in which we measured \wprp. We model the
projected correlation functions using HOD formalism, within a flat
$\Lambda \rm{CDM}$ cosmology with parameters identical to those used
in the MultiDark simulation (see Section \ref{sec:mocks}). We restrict
the fit to scales above $\smash{r_p=0.2\mhmpc}$ and below
$r_p=30\mhmpc$ and correct empirically the measured projected
correlation function for the residual underestimation at different
scales, using the ratio between the parent and recovered \wprp in the
observed mocks for the same galaxy selection. We assume that there is
negligible error in taking this small correction to be independent of
cosmology. In the fitting procedure we used both the sample number
density and \wprp constraints in order to estimate the HOD parameters
and their errors by exploring the full parameter space of the model.

In our HOD model the occupation number is parameterized as
\begin{equation}
  \left<N_{\rm gal}|m\right>=\left<N_{\rm cen}|m\right>(1+\left<N_{\rm sat}|m\right>) \label{eq:HOD}
\end{equation}
where $\left<N_{\rm cen}|m\right>$ and $\left<N_{\rm sat}|m\right>$
are the average number of central and satellite galaxies in a halo of
mass $m$. This model explicitly assumes that the first galaxy in
haloes, when haloes have reached a sufficient mass, has to be
central. Central and satellite galaxy occupations are defined as in
\citep{zheng05}:
\begin{align}
  \left<N_{\rm cen}|m\right> &= \frac{1}{2}\left[1+\rm{erf}\left(\frac{\log~m - \log
      M_{\rm min}}{\sigma_{\log~m}}\right)\right], \label{ncen}  \\
  \rlap{$\left<N_{\rm sat}|m\right>$}
  \phantom{\left<N_{\rm cen}|m\right>}
&= \left(\frac{m-M_0}{M_1}\right)^{\alpha}. 
\label{nsat}
\end{align}
where $M_{\rm min}$, $\sigma_{\log~m}$, $M_{0}$, $M_{1}$, and $\alpha$
are the HOD parameters. The parameter $M_{0}$ is generally poorly
constrained and we decided in this analysis to fix $M_{0}=M_{\rm min}$
\citep[see also][]{white11,delatorre12}.

In the halo model formalism, the galaxy power spectrum or two-point
correlation function can be written as the sum of two component: the
1-halo term that describes the correlations of galaxies inside haloes
and the 2-halo term that characterises the correlations of galaxies
sitting in different halos. We follow the formalism of
\citet{vandenbosch13} to define the projected correlation in the
context of this model. In particular we use their improved
prescriptions for the treatment of the halo-exclusion and residual
redshift-space distortions effects on \wprp, induced by the finite
$\pi_{\rm max}$ values used in the data \citep{vandenbosch13}. We use
the halo bias factor and mass function of \citet{tinker08} and
\citet{tinker10} respectively, and assume that satellite galaxies
trace the mass distribution within haloes. We make the assumption of a
NFW \citep{navarro96} radial density profile and use the
concentration-mass relation obtained by \citet{prada12} from the
MultiDark simulation. The details of the implementation of the HOD
model are given in de la Torre at al. (in preparation).

We present in Fig. \ref{fig12} the measurements and best-fitting HOD
models for the four different volume-limited absolute
magnitude-threshold samples. We find that the model reproduces the
observations well. To have a global characterisation of the clustering
properties of galaxies in VIPERS, we extend this modelling to two
additional redshift bins at $0.5<z<0.7$ and $0.9<z<1.1$. The
best-fitting $M_{\rm min}$ and $M_{1}$ parameters for the different
sub-samples are shown in Fig. \ref{fig13} and compared to previous
measurement in the same range of redshift and number density. Because
in the different surveys the subsamples are not selected with the same
absolute magnitude band of selection, it is convenient to compare the
HOD parameters in terms of redshift and the number density probed by
each sample. Note that here we compare measurements only from analyses
using the same HOD parameterization, although the exact implementation
of the models can differ slightly. The VIPERS sample allows us to
constraint these parameters with an unprecedented accuracy over the
redshift range $0.5<z<1.1$. Our results are consistent with previous
measurements, in particular with the DEEP2 \citep{zheng07} and CFHTLS
\citep{coupon12} analyses. Our HOD analysis is aimed at modelling the
global clustering properties in VIPERS, but we refer the reader to
\citet{marulli13} and de la Torre et al. (2013, in preparation) for
detailed analysis and interpretation of the luminosity and stellar
dependence of galaxy clustering and luminosity-dependent halo
occupation respectively.

\begin{figure}
\resizebox{\hsize}{!}{\includegraphics{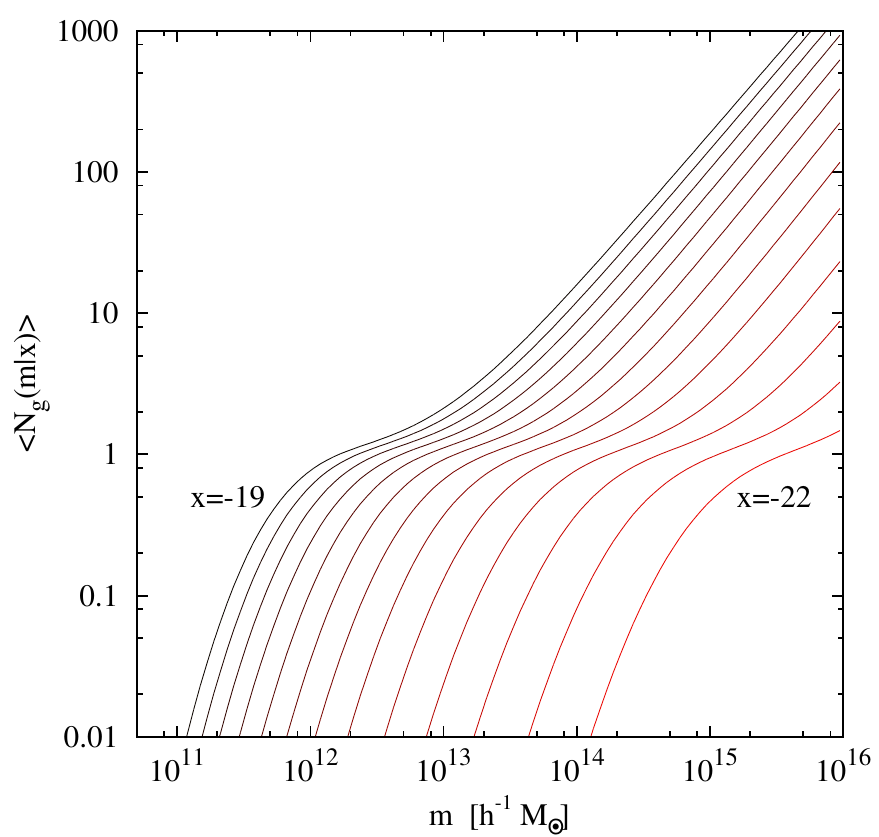}}
\caption{Evolution of the B-band absolute magnitude-dependent HOD. The
  curves show $\langle N_{\rm gal}(m|z,M_B) \rangle$ for values of
  $x=M_B-5\log(h)+z$ ranging from $x=-19$ to $x=-22$ with steps of
  $\Delta x=0.25$, respectively from left to right.}
\label{fig14}
\end{figure}

We use the derived HOD parameters to define a global luminosity- and
redshift-dependent occupation number which is then used to create
accurate HOD mocks of the survey. In order to interpolate between the
different redshifts we assume a global luminosity evolution
proportional to redshift, so that the magnitude threshold values scale
linearly with redshift \citep{brown08,coupon12}. We find that one can
approximate $\langle N_{\rm gal}(m|z,M_B) \rangle$ using Eq.
\ref{eq:HOD} with
\begin{align}
\log M_{\rm min}(x) &= 10.61\exp(1.49^{-24.66-x}) \\
\sigma_{\log m}(x) &= 0.06\exp(-0.08x+0.34) \\
M_0(x) &= M_{\rm min}(x) \\
M_1(x) &= 13.5 M_{\rm min}(x) \\
\alpha(x) &= 0.29\exp(-0.05x+0.38)\, ,
\end{align}
where $x=M_B-5\log(h)+z$. $M_{\rm min}$ and $M_{1}$ are found to be
strongly correlated in such a way that $M_{1}$ is approximately equal
to $10-20$ times $M_{\rm min}$ depending on the redshift probed and
the model implementation \citep[e.g.][]{beutler13}. In our analysis we
find that $M_1(x)$ can be approximated by $13.5$ times $M_{\rm
  min}(x)$. The function $\langle N_{\rm gal}(m|z,M_B) \rangle$ is
shown in Fig. \ref{fig14} for the different values of $x$ probed with
VIPERS.  We checked the consistency of this parameterization and
verify that the \wprp predicted by the mocks and the that measured are
in good agreement for all probed redshift and luminosity thresholds.

\section{Redshift-space distortions}

The main goal of VIPERS is to provide with the final sample accurate
measurements of the growth rate of structure in two redshift bins
between $z=0.5$ and $z=1.2$. The growth rate of structure $f$ can be
measured from the anisotropies observed in redshift space in the
galaxy correlation function or power spectrum. Although this
measurement is degenerate with galaxy bias, the combination
$f\sigma_8$ is measurable and still allows a fundamental test of
modifications of gravity since it is a mixture of the differential and
integral growth. In this Section, we present an initial measurement of
$f\sigma_8$ from the VIPERS first data release.

\subsection{Method}

With the first epoch VIPERS data we can reliably probe scales below
about 35\hmpc. The use of the smallest non-linear scales,
i.e. typically below $10\mhmpc$, is however difficult because of the
limitations of current redshift-space distortion models, which cannot
describe the non-linear effects that relate the evolution of density
and velocity perturbations. However, with the recent developments in
perturbation theory and non-linear models for RSD
\citep[e.g.][]{taruya10,reid11,seljak11}, we can push our analysis
well into mildly non-linear scales and obtain unbiased measurements of
$f\sigma_8$ while considering minimum scales of $5-10\mhmpc$
\citep{delatorre12}.

\begin{figure}
\resizebox{\hsize}{!}{\includegraphics{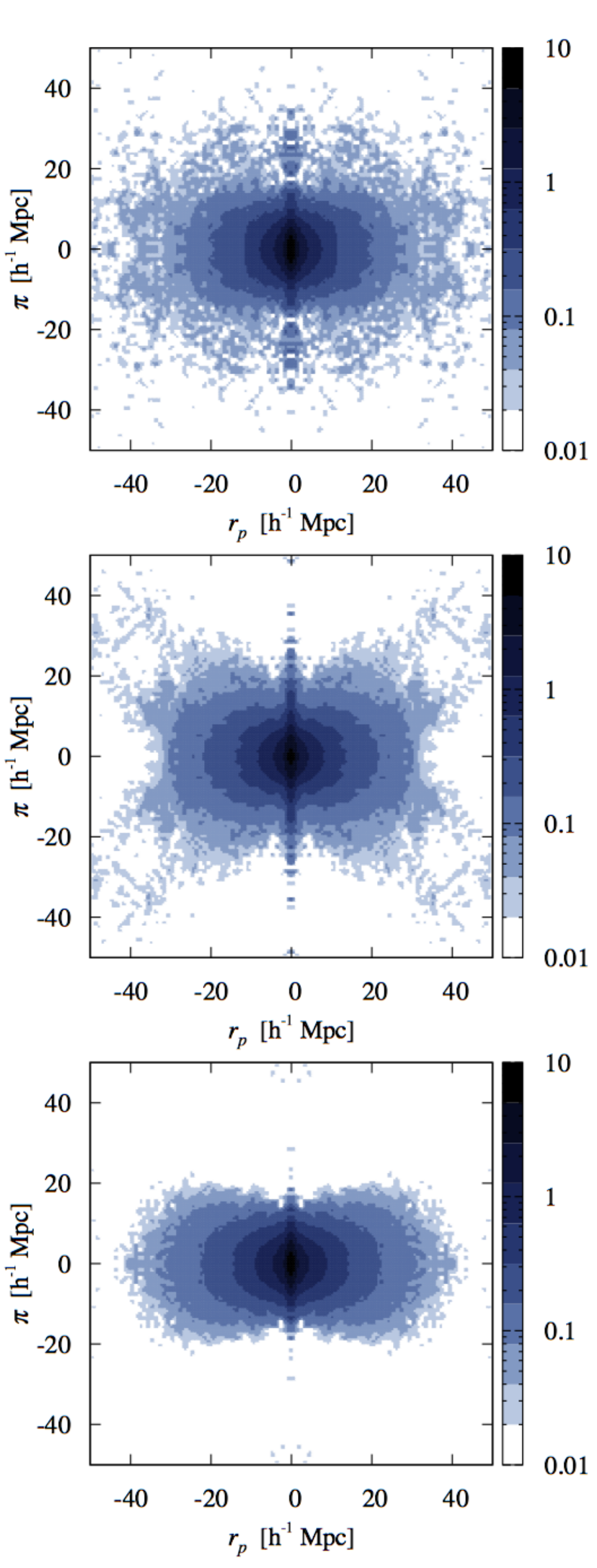}}
\caption{Anisotropic correlation functions of galaxies at
  $0.7<z<1.2$. The top panel shows the results for the VIPERS first
  data release, deduced by the Landy-Szalay estimator counting pairs
  in cells of side $1\mhmpc$. The lower two panels show the results of
  two simulations, which span the 68\% confidence range on the fitted
  value of the large-scale flattening (see Section
  \ref{sec:rsdfits}).}
\label{fig15}
\end{figure}

With the VIPERS first data release, we perform an initial
redshift-space distortion analysis, considering a single redshift
interval of $0.7<z<1.2$ to probe the highest redshifts where the
  growth rate is little-konwn. We select all galaxies above the
magnitude limit of the survey in that interval. The effective
pair-weighted mean redshift of the subsample is $z=0.80$. The measured
anisotropic correlation function \xisp is shown in the top panel of
Fig. \ref{fig15}. We have used here a linear binning of $\smash{\Delta
  r_p=\Delta \pi=1\mhmpc}$. One can see in this figure the two main
redshift-space distortion effects: the elongation along the
line-of-sight, or Finger-of-God effect, which is due to galaxy random
motions within virialized objects and the squashing effect on large
scales, or Kaiser effect, which represents the coherent large-scale
motions of galaxies towards overdensities. The latter effect is the
one we are interested in since its amplitude is directly related to
the growth rate of perturbations. Compared to the first
  measurement at such high redshift done with the VVDS survey
  \citep{guzzo08}, this signature is detected with much larger
  significance, with the flattening being apparent to $r_p>30\mhmpc$.

The anisotropic correlation has been extensively used in the
literature to measure the growth rate or the distortion parameter
$\beta$
\citep[e.g.][]{hawkins03,guzzo08,cabre09,beutler12,contreras13}. However,
with the increasing size and statistical power of redshift surveys, an
alternative approach has grown in importance: the use of the multipole
moments of the anisotropic correlation function. This approach has the
main advantage of reducing the number of observables, compressing the
cosmological information contained in the correlation function. In
turn, this eases the estimation of the covariance matrices associated
with the data. We adopt this methodology in this analysis and fit for
the two first non-null moments $\xi_0(s)$ and $\xi_2(s)$, where most
of the relevant information is contained, and ignore the contributions
of the more noisy subsequent orders. The multipoles moments are
measured from $\xi(s,\mu)$ which is obtained exactly as for \xisp,
except that the redshift-space separation vector $\vec{s}$ is now
decomposed into the polar coordinates $(s,\mu)$ such that
$r_p=s(1-\mu^2)^{1/2}$ and $\pi=s\mu$. The multipole moments are
related to $\xi(s,\mu)$ as,
\begin{equation}
\xi_\ell(s)=\frac{2\ell+1}{2}\int_{-1}^{1}\xi(s,\mu)L_\ell(\mu)d\mu, \label{eq:xil}
\end{equation}
where $L_\ell$ is the Legendre polynomial of order $\ell$. In practice the
integration of Eq. \ref{eq:xil} is approximated by a Riemann sum over
the binned $\xi(s,\mu)$. We use a logarithmic binning in $s$ of
$\Delta \log(s)=0.1$ and linear binning in $\mu$ with $\Delta
\mu=0.02$.

\begin{figure}
\resizebox{\hsize}{!}{\includegraphics{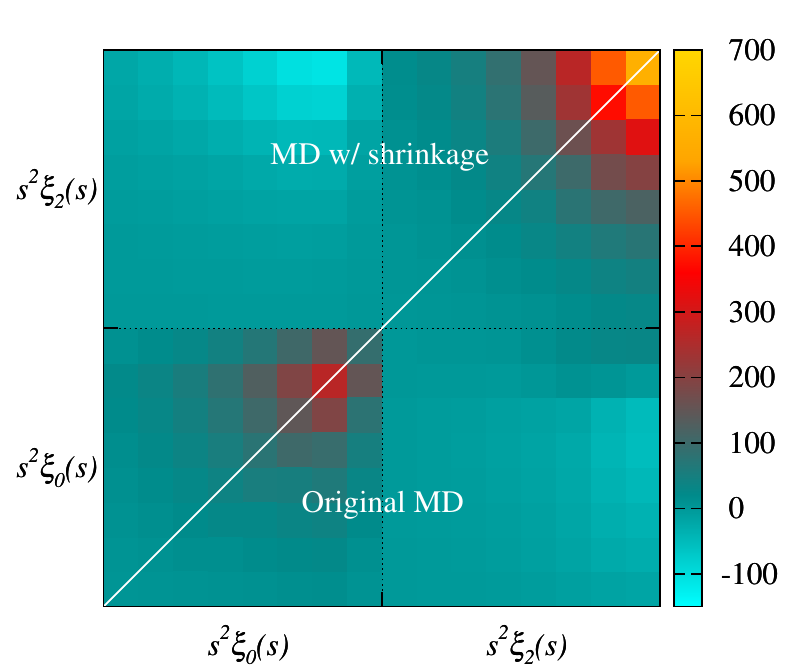}}
\caption{Covariance matrix for the redshift-space distortions
  analysis. While the lower triangle represents the covariance matrix
  obtained directly from the MD mocks, the upper triangle corresponds
  to that obtained after performing shrinkage estimation.}
\label{fig16}
\end{figure}

\subsection{Covariance matrix, error estimation, and fitting procedure}

The different bins in the observed correlation function and associated
multipole moments are correlated to some degree, and this must be
allowed for in order to fit the measurements with theoretical
models. We estimate the covariance matrix of the monopole and
quadrupole signal using the MultiDark (MD) and Pinocchio (PN) HOD
mocks. The generic elements of the matrix can be evaluated as
\begin{equation}
C_{ij}=\frac{1}{N_{R}-1}\sum_{k=1}^{N_R}\left(y^k(s_i)-\bar{y}(s_i)\right)\left(y^k(s_j)-\bar{y}(s_j)\right)
\end{equation}
where $N_R$ is the number of mock realizations, $y(s)$ is the quantity
of interest, and the indices $i,j$ run over the data points.

The number of degrees of freedom in the multipole moments varies
between $11$ and $15$ depending on the scales considered. Because we
have only 26 MD mock realizations, the covariance matrix elements
cannot be constrained accurately with the MD mocks only: the
covariance matrix is unbiased, but it can have substantial noise. In
order to mitigate the noise and obtain an accurate estimate of the
covariance matrix, we apply the shrinkage method \citep{pope08}, using
the covariance matrix obtained with the 200 PN mocks as the target
matrix. The PN mocks are more numerous and therefore each element of
the associated covariance matrix is very well constrained, although
the covariance may be biased to some extent. This bias is related to
inaccuracies in the predicted moments, which are mainly driven by the
limited accuracy of the Zel'dovich approximation used in the PN mocks
to predict the peculiar velocity field. The shrinkage technique allows
the optimal combination of an empirical estimate of the covariance
with a target covariance, minimising the total mean squared error
compared to the true underlying covariance. An optimal covariance
matrix $C$ is then obtained with
\begin{equation}
C=\lambda T + (1-\lambda) S,
\end{equation}
where $\lambda$ is the shrinkage intensity and the target ${T}$ and
empirical ${S}$ covariance matrices correspond respectively to those
obtained from the PN and MD mocks. $\lambda$ is calculated from
\citep{pope08}
\begin{equation} \label{shrink}
\lambda=\frac{\sum_{i,j} {\rm Cov}(S_{ij},S_{ij})-{\rm Cov}(T_{ij},S_{ij})}{\sum_{i,j} ({T}_{ij}-{S}_{ij})^2}  ,
\end{equation}
where ${\rm Cov}(A_{ij},B_{ij})$ stands for the covariance between the
elements $(i,j)$ of the matrices $A$ and $B$. We note that, since the
empirical and target matrices are independent, the term ${\rm
  Cov}(T_{ij},S_{ij})$ vanishes in the numerator of
Eq. \ref{shrink}. The effect of shrinkage estimation on the MD
covariance matrix is shown in Fig. \ref{fig16}.

To measure the growth rate of structure we perform a maximum
likelihood analysis of the data given models of redshift-space
distortions by adopting the likelihood function $\mathcal{L}$:
\begin{equation}
  -2\ln{\mathcal{L}}=\sum_{i=1}^{N_p}\sum_{j=1}^{N_p}\Delta_i C^{-1}_{ij} \Delta_j,
\end{equation}
where $N_p$ is the number of points in the fit, $\Delta$ is the
data-model difference vector, and $C$ is the covariance matrix. The
likelihood is performed on the quantity $y(s)=s^2\xi_\ell(s)$, rather
than simply $y(s)=\xi_\ell(s)$ to reduce the range of variations of
multipole values at different $s$ in the fit. In the end, the quantity
which is matched with model predictions is the concatenation of
$s^2\xi_0$ and $s^2\xi_2$ for the set of separations considered.

As a final remark, we note that we use the direct inverse of the
covariance matrix without applying the correction discussed by
\citet{hartlap07}, as it is not clear how the size of the correction
is affected by the shrinkage estimation technique. The resulting
errors derived from the likelihood are well matched to the
distribution of best-fit values from the mocks, which gives us
confidence that only a small correction, if any, would be necessary.

\begin{figure}
\resizebox{\hsize}{!}{\includegraphics{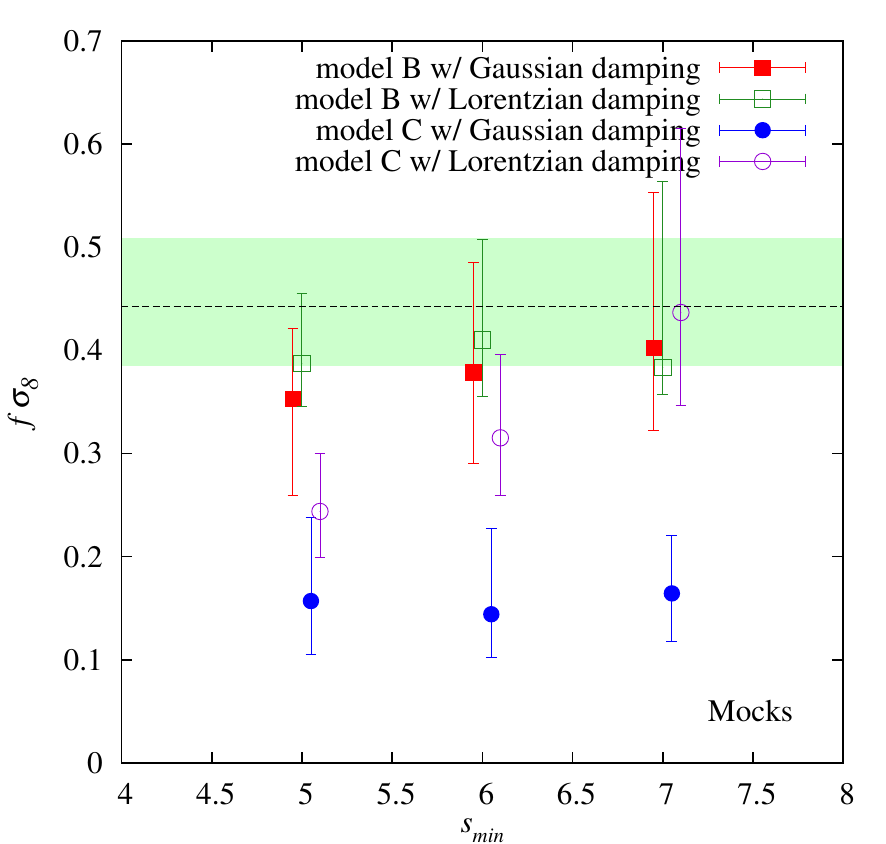}}
\caption{Systematic and statistical errors on $f\sigma_8$ obtained
  from the mock samples, for different models and minimum scales
  $s_{min}$ in the fit. The horizontal dashed line shows the
  expectation value while the shaded area marks the $15\%$ region
  around the fiducial value. Note that the points have been slightly
  displaced horizontally around $s_{min}$ values of $5\mhmpc$,
  $6\mhmpc$, and $7\mhmpc$ to improve the clarity of the figure.}
\label{fig17}
\end{figure}

\subsection{Models}

The formalism that describes redshift-space anisotropies in the power
spectrum can be derived from writing the mass density conservation in
real and redshift space \citep{kaiser87}. In particular, in the
plane-parallel approximation, which is assumed in this analysis, the
anisotropic power spectrum of mass has the general compact form
\citep{scoccimarro99}
\begin{align}
P^s(k,\nu)&={}\int \frac{d^3\vec{r}}{(2\pi)^3} e^{-i\vec{k} \cdot \vec{r}}\left<e^{-ikf\nu \Delta u_\parallel} \times \right. \nonumber \\ 
& \left. [\delta(\vec{x})+\nu^2 f \theta(\vec{x})][\delta(\vec{x}^\prime)+\nu^2 f \theta(\vec{x}^\prime)]\right> \label{eq:rspk}
\end{align}
where $\nu=k_\parallel/k$,
$u_\parallel(\vec{r})=-v_\parallel(\vec{r})/(f aH(a))$,
$v_\parallel(\vec{r})$ is the line-of-sight component of the peculiar
velocity, $\delta$ is the density field, $\theta$ is the divergence of
the velocity field, $\Delta
u_\parallel=u_\parallel(\vec{x})-u_\parallel(\vec{x}^\prime)$ and
$\vec{r}=\vec{x}-\vec{x}^\prime$. Although exact, Eq. \ref{eq:rspk} is
impractical for direct use on redshift survey measurements and several
models have been proposed to approximate it. In the assumption that
galaxies linearly trace the underlying mass density field with a bias
$b$, we can build three empirical models. These take the form,
\begin{equation}
P_g^s(k,\nu)=D(k\nu\sigma_v)P_K(k,\nu;f,b)  \label{eq:models}
\end{equation}
where,
\begin{align}
D(k\nu\sigma_v)=\left\{ \nonumber
\begin{array}{lcl}
\exp(-(k\nu\sigma_v)^2)
\\
\\
1/(1+(k\nu\sigma_v)^2)
\end{array}
\right.
\end{align}
and,
\begin{align}
P_K(k,\nu;f,b) = \hspace{7.0cm} & \nonumber \\ 
\left\{ \nonumber
\begin{array}{lcl}
\rlap{$b^2 \Pdd(k)+2\nu^2 fb \Pdd(k) +\nu^4 f^2 \Pdd(k)$} \hspace{5.5cm} {\rm (model~A)} \\
\\
\rlap{$b^2 \Pdd(k)+2\nu^2 fb \Pdt(k) +\nu^4 f^2 \Ptt(k)$} \hspace{5.5cm} {\rm (model~B)}  \\
\\
b^2 \Pdd(k)+2\nu^2 fb \Pdt(k) +\nu^4 f^2 \Ptt(k) \\ 
\rlap{$+ C_A(k,\nu;f,b) + C_B(k,\nu;f,b).$} \hspace{5.5cm} {\rm (model~C)} 
\end{array}
\right.
\end{align}
In these equations $\Pdd$, $\Pdt$, $\Ptt$ are respectively the
non-linear mass density-density, density-velocity divergence, and
velocity divergence-velocity divergence power spectra and $\sigma_v$
is an effective pairwise velocity dispersion that we can fit for and
then treat as a nuisance parameter. The expressions for
$C_A(k,\nu;f,b)$ and $C_B(k,\nu;f,b)$ are given in appendix A of
\citet{delatorre12}. These empirical models can be seen in
configuration space, as a convolution of a damping function
$D(k\mu\sigma_v)$, which we assume to be Gaussian or Lorentzian in
Fourier space, and a term involving the density and velocity
divergence correlation functions and their spherical Bessel
transforms. While the first term essentially (but not only) describes
the Finger-of-God effect, the second, $P_K(k,\nu,b)$, describes the
Kaiser effect. We note that model A is the classical dispersion model
\citep{peacock94} based on the linear \citet{kaiser87} model; model B
is the generalisation proposed by \citet{scoccimarro04} that accounts
for the non-linear coupling between the density and velocity fields,
making explicitly appearing the velocity divergence auto-power
spectrum and density--velocity divergence cross-power spectrum; model
C is an extension of model B that contains the two additional
correction terms proposed by \citet{taruya10} to correctly account for
the coupling between the Kaiser and damping terms. We refer the reader
to \citet{delatorre12} for a thorough description of these models.

In the end, the model $\xi^s_\ell(s)$ are obtained from their Fourier
counterparts,
\begin{equation} \label{expmomK}
P^s_\ell(k)=\frac{2\ell+1}{2} \int_{-1}^1 d\nu P_g^s(k,\nu) L_\ell(\nu),
\end{equation}
as
\begin{equation} \label{expmom}
\xi^s_\ell(s)=i^\ell \int \frac{dk}{2\pi^2} k^2 P^s_\ell(k)j_\ell(ks),
\end{equation} 
where $j_\ell$ denotes the spherical Bessel functions.

The redshift-space distortion models involve the knowledge of the
underlying mass non-linear power spectra of density and velocity
divergence at the effective redshift of the sample. Although the
real-space non-linear correlation function of galaxies can be
recovered from the deprojection of the observed projected correlation
function \citep[e.g.][]{bianchi12} and thus be used to some extent as
an input of the model \citep[e.g.][]{hamilton92}, it is not feasible
for the more advanced models which involve the velocity divergence
power spectrum. The non-linear power spectra can however be predicted
from perturbation theory or simulations for different cosmological
models. In the case of a $\Lambda \rm{CDM}$ cosmology, the shape of
the non-linear power spectra depends on the parameters $P=(\Omega_m,
\Omega_b, h, n_s, \sigma_8)$ and can be obtained to a great accuracy
from semi-analytical prescriptions such as {\sc HALOFIT}. In this
analysis we use the latest calibration of {\sc HALOFIT} by
\citet{bird12} to obtain $\Pdd$ and use the fitting functions of
\citet{jennings12} to predict $\Ptt$ and $\Pdt$ from $\Pdd$. The
latter fitting functions are accurate at the few percent level up to
$k\simeq0.3$ at $z=1$.

In the models, the bias and growth rate parameters $b$ and $f$ are
degenerate with the normalization of the power spectrum parameter
$\sigma_8$. Thus, in practice only the combination of $b\sigma_8$ and
$f\sigma_8$ can be constrained if no assumption is made on the actual
value of $\sigma_8$. This can be done by renormalizing the power
spectra in the models so that $P_{xx}(k,z) \rightarrow
P_{xx}(k,z)/\sigma_8^2(z)$, thus reparameterizing the models such that
the parameters $(b,f)$ are replaced by $(b\sigma_8,f\sigma_8)$ in
Eq. \ref{eq:models}. This parameterization can be used for model A and
B, although not for model C in which the correction term $C_A$
involves the additional combinations: $b^2f\sigma_8^4$,
$bf^2\sigma_8^4$, and $f^3\sigma_8^4$
\citep[see][]{taruya10,delatorre12}. The correction term $C_A$, which
partially describes the effects of the non-linear coupling between the
damping and Kaiser terms, mostly affects the monopole and quadrupole
moments of the redshift-space power spectrum on scales of $k>0.1$
\citep{taruya10}. Therefore, in principle $C_A$ could help breaking
the degeneracy between $f$ and $\sigma_8$ although this has to be
verified in detail. In the end, in the case of model C we decided to
treat $(f,b,\sigma_8,\sigma_v)$ as separate parameters in the fit.

\subsection{Detailed tests against mock data}\label{sec:rsdfits}

\begin{figure}
\resizebox{\hsize}{!}{\includegraphics{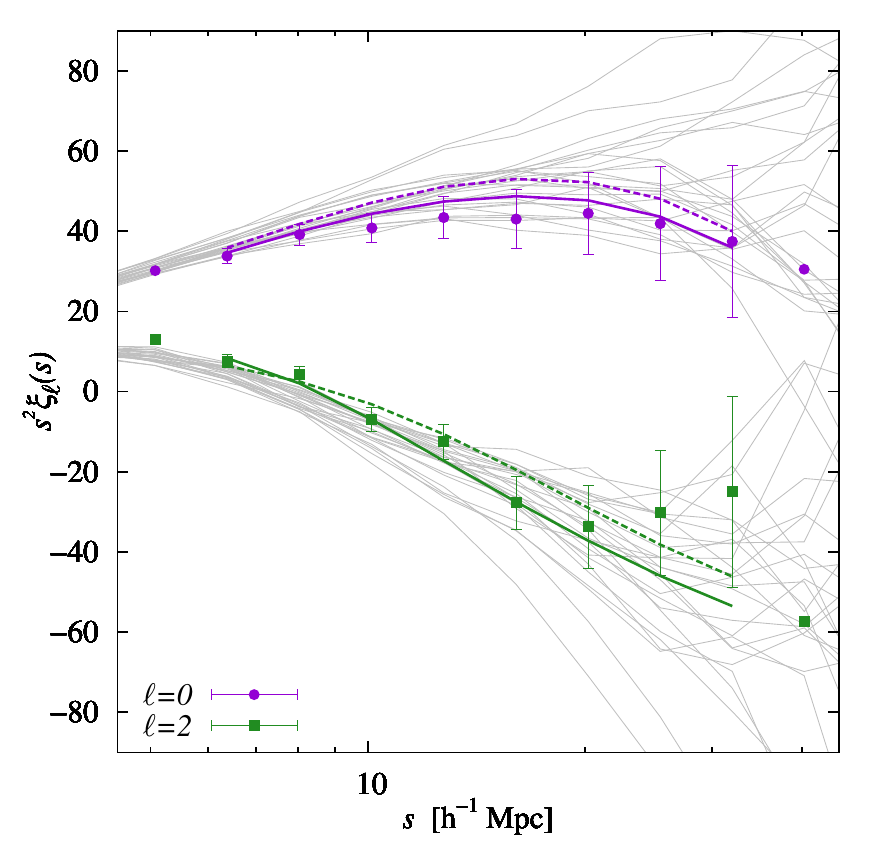}}
\caption{Monopole and quadrupole moments of the redshift-space
  correlations, as a function of scale. The shallow curves show the
  results for the 26 individual MultiDark simulation mocks; the points
  are for the measured VIPERS data at $0.7<z<1.2$, with assigned error
  bars based on the scatter in the mocks. The solid and dotted lines
  correspond to the best fitting models to the data for model B with
  Gaussian or Lorentzian damping function respectively.}
\label{fig18}
\end{figure}

\begin{figure}
\resizebox{\hsize}{!}{\includegraphics{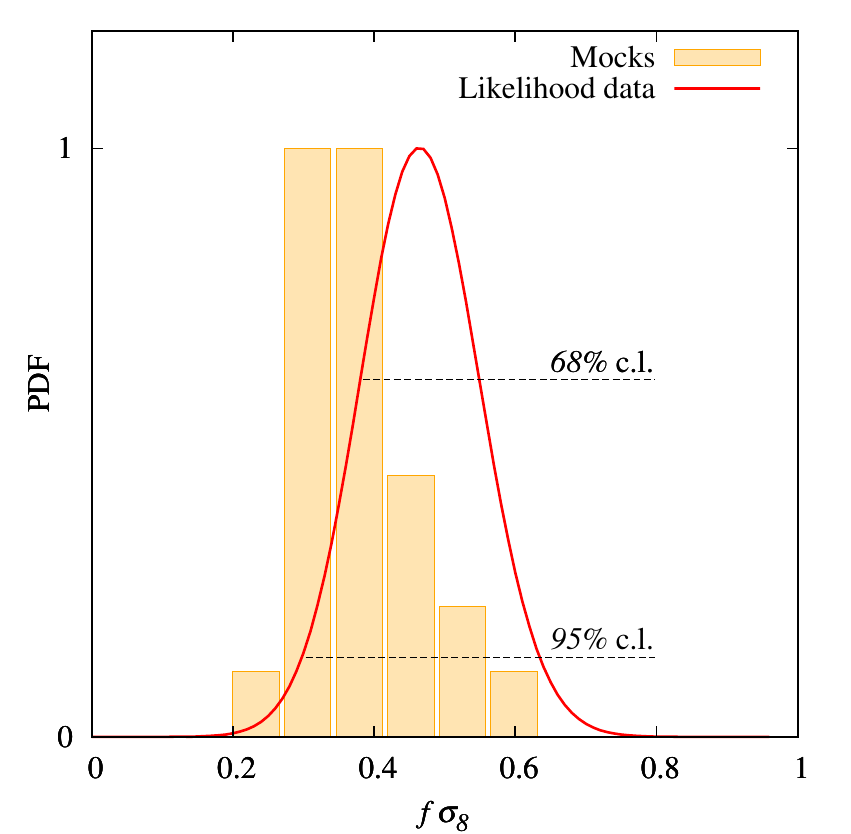}}
\caption{Marginalized likelihood distribution of $f\sigma_8$ in the
  data (solid curve) and distribution of fitted values of $f\sigma_8$
  for the 26 individual MultiDark simulation mocks (histogram). These
  curves show a preferred value and a dispersion in the data that is
  consistent at the $1\sigma$ level with the distribution over the
  mocks.}
\label{fig19}
\end{figure}

\begin{figure*}[ht]
\centering
\includegraphics[width=13cm]{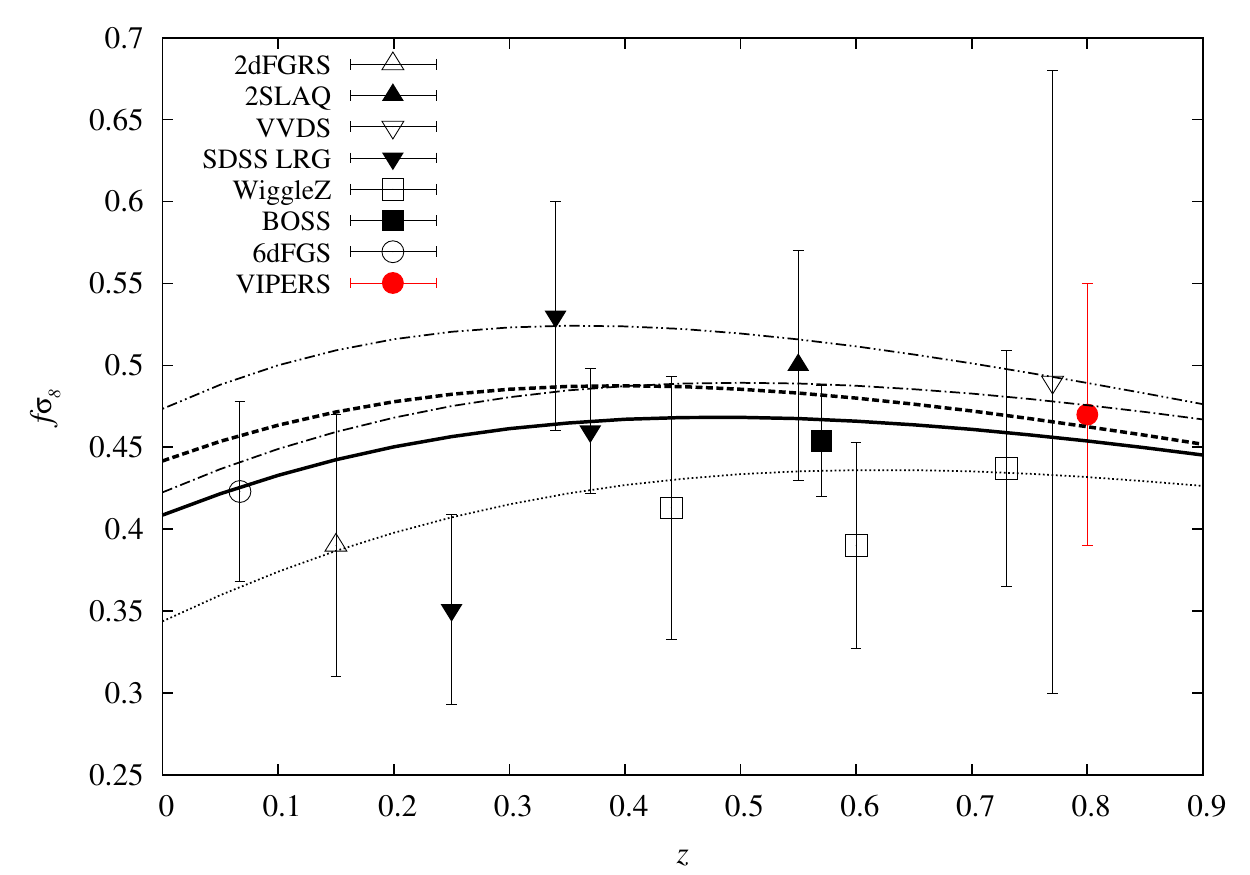}
\caption{A plot of $f\sigma_8$ versus redshift, showing VIPERS result
  contrasted with a compilation of recent measurements. The previous
  results from 2dFGRS \citep{hawkins03}, 2SLAQ \citep{ross07}, VVDS
  \citep{guzzo08}, SDSS LRG \citep{cabre09,samushia12}, WiggleZ
  \citep{blake12}, BOSS \citep{reid12}, and 6dFGS \citep{beutler12}
  surveys are shown with the different symbols (see inset). The thick
  solid (dashed) curve corresponds to the prediction for General
  Relativity in a $\Lambda \rm{CDM}$ model with WMAP9 (Planck)
  parameters, while the dotted, dot-dashed, and dot-dot-dashed curves
  are respectively Dvali-Gabadaze-Porrati \citep{dvali00}, coupled
  dark energy, and $f(R)$ model expectations. For these models, the
  analytical growth rate predictions given in \citet{diporto12} have
  been used.}
\label{fig20}
\end{figure*}

We perform the redshift-space distortion analysis of the VIPERS data
in the context of a flat $\Lambda\rm{CDM}$ cosmological model. Before
considering the redshift-space distortions in the data, we first test
the methodology and expected errors on $f\sigma_8$ using the mock
samples. We fix the shape of the mass non-linear power spectrum to
that of the simulation (since the observed real-space correlations are
of high accuracy) and perform a likelihood analysis of each individual
MD mock. In the case of model C we also fix the normalisation of the
power spectrum as discussed above. The distribution of best-fitting
$f\sigma_8$ values gives us a direct estimate of the probability
distribution function of the parameter for a given fitting method, and
serves as a check on the errors from the full likelihood function.  We
estimate the median and $68\%$ confidence region of the
distribution. These are shown in Fig. \ref{fig17} for the different
models presented in the previous section and for various minimum
scales $s_{min}$ in the fit.

Model A is known to be the most biased model
\citep[e.g.][]{okumura11,bianchi12,delatorre12} and our results
confirm these findings. We thus decided not to describe in the
following the detailed behaviour of this model and focus on models B
and C. We find that in general model B tends to be less biased than
model C, which is surprising at first sight as model C is the most
advanced and supposed to be the most accurate
\citep{kwan12,delatorre12}. This could be due to the quite restricted
scales that we consider and the limited validity of its implementation
on scales below $s\simeq10\mhmpc$, as the maximum wavenumber to which
we can predict $\Pdt$ and $\Ptt$ is about $k=0.3$. We defer the
investigation of this issue to the redshift-space distortion analysis
of the final sample and concentrate here on model B. The shape of the
damping function in the models also affects the recovered $f\sigma_8$,
as expected given the minimum scales we consider, although in the case
of model B the change in $f\sigma_8$ is at most $5\%$.  Including
smaller scales in the fit reduces the statistical error but at the
price of slightly larger systematic error. Therefore from this test we
decided to use model B and a compromise value for the minimum scale of
$s_{min}=6\mhmpc$.

\subsection{The VIPERS result for the growth rate}

These comprehensive tests of our methodology give us confidence that
we can now proceed to the analysis of the real VIPERS data and expect
to achieve results for the growth rate that are robust, and which can
be used as a trustworthy test of the nature of gravity at high
redshifts.

As explained earlier, we assume a fixed shape of the mass power
spectrum consistent with the cosmological parameters obtained from
WMAP9 \citep{hinshaw12} and perform a maximum likelihood analysis on
the data, considering variations in the parameters that are not well
determined externally.  The best-fitting models are shown in
Fig. \ref{fig18} when considering either a Gaussian or a Lorentzian
damping function. Although the mock samples tend to slightly prefer
models with Lorentzian damping as seen in Fig. \ref{fig17}, we find
that the Gaussian damping provides a much better fit to the real data
and we decided to quote the corresponding $f\sigma_8$ as our final
measurement.

We measure a value of
\begin{equation}
f(z=0.8)\sigma_8(z=0.8)=0.47\pm0.08,
\end{equation}
which is consistent with the General Relativity prediction in a flat
$\Lambda \rm{CDM}$ Universe with cosmological parameters given by
WMAP9, for which the expected value is $f(0.8)\sigma_8(0.8)=0.45$. We
find that our result is not significantly altered if we adopt a Planck
cosmology \citep{plank13} for the shape of the mass power spectrum,
changing our best-fitting $f\sigma_8$ by only $0.2\%$. This shows that
given the volume probed by the survey, we are relatively insensitive
to the additional Alcock-Paczynski distortions \citep{alcock79} on the
correlation function. The marginalised likelihood distribution of
$f\sigma_8$ is shown superimposed on the mock results in
Fig. \ref{fig19}. We see that the preferred values of the growth rate
are consistent with the mocks, in terms of the width of the likelihood
function being comparable to the scatter in mock fitted values. To
illustrate the degree of flattening of the anisotropic correlation
function induced by structure growth, we show in the middle and bottom
panels of Fig. \ref{fig15} \xisp for two MD mocks for which the
measured $f\sigma_8$ roughly coincide with the $1\sigma$ limits around
the best-fit $f\sigma_8$ value obtained in the data. We therefore
conclude that the initial VIPERS data prefer a growth rate that is
fully consistent with predictions based on standard gravity. Our
measurement of $f\sigma_8$ is also in good agreement with previous
measurements at lower redshifts from 2dFGRS \citep{hawkins03}, 2SLAQ
\citep{ross07}, VVDS \citep{guzzo08}, SDSS LRG
\citep{cabre09,samushia12}, WiggleZ \citep{blake12}, BOSS
\citep{reid12}, and 6dFGS \citep{beutler12} surveys as shown in
Fig. \ref{fig20}. In particular, it is compatible within $1\sigma$
with the results obtained in the VVDS \citep{guzzo08} and WiggleZ
\citep{blake12} surveys at a similar redshift, although WiggleZ
measurements tend to suggest lower $f\sigma_8$ values, smaller than
expected in standard gravity \citep[but see][]{contreras13}.

Finally we compare our measurement to the predictions of three of the
most plausible modified gravity models studied in
\citet{diporto12}. We consider Dvali-Gabadaze-Porrati
\citep[DGP,][]{dvali00}, $f(R)$, and coupled dark energy models and
show their predictions in Fig. \ref{fig20} \citep[see][for the detail
  of their analytic predictions]{diporto12}. We find that our
$f\sigma_8$ measurement is currently unable to discriminate between
these modified gravity models and standard gravity given the size of
the uncertainty, although we expect to improve the constraints with
the analysis of the VIPERS final dataset.

\section{Conclusions}

We have analysed in this paper the global real- and redshift-space
clustering properties of galaxies in the VIPERS survey first data
release. We have presented the selection function of the survey and
the corrections that are needed in order to derive estimates of galaxy
clustering that are free of observational biases. This has been
achieved by using a large set of simulated mock realizations of the
survey to quantify in detail the systematics and uncertainties on our
clustering measurements.

The first data release of about $54000$ galaxies at $0.5<z<1.2$ in the
VIPERS survey allows a measurement of the real-space clustering of
galaxies through the measurement of the projected two-point
correlation function, to an unprecedented accuracy over
$0.5<z<1.2$. This permits detailed modelling of the halo occupation
distribution at these redshifts to be carried out. From an initial HOD
modelling of $B$-band luminosity selected samples, we have been able
to accurately determine the characteristic halo masses for halo
occupation in the redshift interval $0.5<z<1.0$. These measurements
are invaluable for creating realistic synthetic mock samples.

The main goal of VIPERS is to provide an accurate measurement of the
growth rate of structure through the characterisation of the
redshift-space distortions in the galaxy clustering pattern. With the
first data release we have been able to provide an initial measurement
of $f\sigma_8$ at $z=0.8$. We find a value of $f\sigma_8=0.47\pm0.08$
which is in agreement with previous measurements at lower
redshifts. This allows us to put a new constraint on gravity at the
epoch when the Universe was almost half its present age. Our
measurement of $f\sigma_8$ is statistically consistent with a Universe
where the gravitational interactions between structures on $10 \mhmpc$
scales can be described by Einstein's theory of gravity.

The present dataset represents the half-way stage of the VIPERS
project, and the final survey will be large enough to subdivide our
measurements and follow the evolution of $f\sigma_8$ out to redshift
one. This will allow us to address some issues such as the suggestion
from the WiggleZ measurements that $f\sigma_8$ is lower than expected
at $z>0.5$. Our measurement at $z=0.8$ already argues against such a
trend to some extent, but the larger redshift baseline and tighter
errors from the final VIPERS dataset can be expected to deliver a
definitive verdict on the the high-redshift evolution of the strength
of gravity.

\begin{acknowledgements}

We acknowledge the crucial contribution of the ESO staff for the
management of service observations. In particular, we are deeply
grateful to M. Hilker for his constant help and support of this
program. Italian participation to VIPERS has been funded by INAF
through PRIN 2008 and 2010 programs. LG acknowledges support of the
European Research Council through the Darklight ERC Advanced Research
Grant (\# 291521). OLF acknowledges support of the European Research
Council through the EARLY ERC Advanced Research Grant (\#
268107). Polish participants have been supported by the Polish
Ministry of Science (grant N N203 51 29 38), the Polish-Swiss Astro
Project (co-financed by a grant from Switzerland, through the Swiss
Contribution to the enlarged European Union), the European Associated
Laboratory Astrophysics Poland-France HECOLS and a Japan Society for
the Promotion of Science (JSPS) Postdoctoral Fellowship for Foreign
Researchers (P11802). GDL acknowledges financial support from the
European Research Council under the European Community's Seventh
Framework Programme (FP7/2007-2013)/ERC grant agreement n. 202781. WJP
and RT acknowledge financial support from the European Research
Council under the European Community's Seventh Framework Programme
(FP7/2007-2013)/ERC grant agreement n. 202686. WJP is also grateful
for support from the UK Science and Technology Facilities Council
through the grant ST/I001204/1. EB, FM and LM acknowledge the support
from grants ASI-INAF I/023/12/0 and PRIN MIUR 2010-2011. PM
acknowledges the support from the grant PRIN MIUR 2010-2011. \\

This work made use of the facilities of HECToR, the UK's national
high-performance computing service, which is provided by UoE HPCx Ltd
at the University of Edinburgh, Cray Inc and NAG Ltd, and funded by
the Office of Science and Technology through EPSRC's High End
Computing Programme. We are grateful to Ken Rice for assisting us 
with accessing HECToR facilities. \\

The MultiDark Database used in this paper and the web application
providing online access to it were constructed as part of the
activities of the German Astrophysical Virtual Observatory as result
of a collaboration between the Leibniz-Institute for Astrophysics
Potsdam (AIP) and the Spanish MultiDark Consolider Project
CSD2009-00064. The Bolshoi and MultiDark simulations were run on the
NASA's Pleiades supercomputer at the NASA Ames Research Center.

\end{acknowledgements}

\bibliographystyle{aa}
\bibliography{biblio}

\appendix

\section{Galaxy mock catalogue construction}

We provide in this appendix some details about the method that we used
to create realistic galaxy catalogues based on the Halo Occupation
Distribution (HOD) and Stellar-to-Halo Mass Relation (SHMR)
formalisms. From the MultiDark simulation and Pinocchio halo
lightcones described in Section \ref{sec:mocks}, we created two types
of galaxy mock catalogues: one containing $B$-band absolute magnitudes
and associated quantities, and a second one containing stellar masses.
We note that the stellar mass mock catalogues have
not been explicitly used in this analysis, but in the accompanying
VIPERS analyses of \citet{marulli13} and \citet{davidzon13}.

For the first set of catalogues we use the HOD formalism and populated
dark matter haloes according to their mass by specifying the absolute
$B$-band magnitude-dependent halo occupation. We parametrised the
latter using Eq. \ref{eq:HOD} and used the HOD parameters obtained
from the data and given in Section \ref{sec:realclus}. We positioned
central galaxies at halo centres with probability given by a Bernoulli
distribution function with mean taken from Eq. \ref{ncen} and assigned
host halo mean velocities to these galaxies. The number of satellite galaxies
per halo is set to follow a Poisson distribution with mean given by
Eq. \ref{nsat}. We assumed that satellite galaxies follow the spatial
and velocity distribution of mass and randomly distributed their
halo-centric radial position so as to reproduce a \citet{navarro96} (NFW)
radial profile,
\begin{equation}
\rho_{NFW}(r|m)\propto\left(\frac{c_{dm}(m)r}{r_v(m)}\right)^{-1}\left(1+\frac{c_{dm}(m)r}{r_v(m)}\right)^{-2},
\end{equation}
where $c_{dm}$ is the concentration parameter and $r_v(m)$ is the
virial radius defined as
\begin{equation}
r_{v}(m)=\left(\frac{3m}{4\pi\bar{\rho}(z)\Delta_{NL}}\right)^{1/3}.
\end{equation}
In this equation, $\bar{\rho}(z)$ is the mean matter density at
redshift $z$ and $\Delta_{NL}=200$ is the critical overdensity for
virialisation in our definition. We assumed the mass-concentration
relation of \citet{bullock01}:
\begin{equation}
c_{dm}(m,z)=\frac{c_0}{1+z}\left(\frac{m}{m_*}\right)^{-0.13},
\end{equation}
where $c_0=11$ and $m_*$ is the non-linear mass scale at $z=0$ defined
such as $\sigma(m_*,0)=\delta_c$. Here $\delta_c$ and $\sigma(m,0)$
are respectively the critical overdensity (we fixed $\delta_c=1.686$)
and the standard deviation of mass fluctuations at $z=0$. The latter
is defined as
\begin{equation}
  \sigma^2(m,z)=\int_0^\infty \frac{dk}{k}
  \frac{k^3P(k,z)}{2\pi^2}|W(kR)|^2 \,\,\,\, ,
\end{equation}
where $R=\left[3m/\left(4\pi\bar{\rho}(z)\right)\right]^{1/3}$,
$P(k,z)$ is the linear mass power spectrum at redshift $z$ in the
adopted cosmology, and $W(x)$ is the Fourier transform of a top-hat
filter. In order to assign satellite galaxy velocities, we assumed
halo isotropy and sphericity, and drew velocities from Gaussian
distribution functions along each Cartesian dimension with velocity
dispersion given by \citep{vandenbosch04}:
\begin{align}
\sigma^2_{sat}(r|m)&={}\frac{1}{\rho_{NFW}(r|m)}\int_r^\infty \rho_{NFW}(r|m)\frac{d\psi}{dr}dr \\ 
&={} \frac{Gm}{r_{v}}\frac{c_{dm}}{f(c_{dm})}\left(\frac{c_{dm} r}{r_v}\right)\left(1+\frac{c_{dm} r}{r_v}\right)^2 I(r/rs),
\end{align}
where $\psi(r)$ is the gravitational potential, $G$ is
the gravitational constant, $f(x)=\ln(1+x)-x/(1+x)$, and
\begin{equation}
  I(x)=\int_x^\infty \frac{f(t)dt}{t^3(1+t)^2}.
\end{equation}

In these mocks, the absolute $B$-band magnitude for each galaxy was
obtained following \citet{skibba06}. From the mean rest-frame $B-i'$
colour and K-corrections observed in the data we then derived absolute
and apparent $i'$-band magnitudes for each simulated galaxy.

For the mock catalogues with stellar masses, we followed the SHMR
approach which is based on the assumption of a monotonic relation
between halo/subhalo masses and the stellar masses of the galaxies
associated with them. We first populated the haloes in the lightcones
with subhaloes. For this we randomly distributed subhaloes around each
distinct halo following a NFW profile so that their number density
satisfies the subhalo mass function of \citet{giocoli10}:
\begin{equation}
\frac{d N(m_{\rm{sub}}|m)}{d \ln m_{\rm{sub}}}= N_0 \xi^\alpha \exp(-\beta\xi^3),
\end{equation}
where $m_{\rm{sub}}$ is the subhalo mass, $\xi=m_{\rm{sub}}/m$,
$\alpha=-0.8$, $\beta=12.2715$, and $N_0=0.18$. We then assigned a
galaxy to each halo and subhalo, with a stellar mass given by the SHMR
of \citet{moster13}. The galaxy velocities were assigned in a similar
way as for the HOD catalogues, with galaxies associated with distinct
haloes and subhaloes being considered as central and satellite
galaxies respectively.

\end{document}